\documentclass[twocolumn]{aastex62}

\usepackage{amsmath}
\usepackage{longtable,booktabs}

\received{}
\revised{}
\accepted{}

\submitjournal{AAS Journals}
\shorttitle{Radii of M Subdwarfs}
\shortauthors{Kesseli et al.}

\begin{document}

\title{Radii of 88 M Subdwarfs and Updated Radius Relations for Low-Metallicity M Dwarf Stars}

\correspondingauthor{Aurora Y. Kesseli}
\email{aurorak@bu.edu}
\author[0000-0002-3239-5989]{Aurora Y. Kesseli}
\affiliation{Department of Astronomy \& Institute for Astrophysical Research, Boston University, 725 Commonwealth Ave., Boston, MA 02215, USA}
\affiliation{IPAC, MS 100-22, Caltech, 1200 E.\ California Blvd., Pasadena, CA 91125, USA}

\author[0000-0003-4269-260X]{J. Davy Kirkpatrick}
\affiliation{IPAC, MS 100-22, Caltech, 1200 E.\ California Blvd., Pasadena, CA 91125, USA}

\author[0000-0001-9309-0102]{Sergio B. Fajardo-Acosta}
\affiliation{IPAC, MS 100-22, Caltech, 1200 E.\ California Blvd., Pasadena, CA 91125, USA}

\author[0000-0001-7506-5640]{Matthew T. Penny}
\affiliation{Department of Astronomy, The Ohio State University, Columbus, Ohio 43210, USA}

\author[0000-0003-0395-9869]{B. Scott Gaudi}
\affiliation{Department of Astronomy, The Ohio State University, Columbus, Ohio 43210, USA}

\author{Mark Veyette}
\affiliation{Department of Astronomy \& Institute for Astrophysical Research, Boston University, 725 Commonwealth Ave., Boston, MA 02215, USA}

\author{Patricia C. Boeshaar}
\affiliation{Department of Physics, University of California, Davis, California 95616, USA}

\author[0000-0001-8877-9060]{Calen B. Henderson}
\affiliation{IPAC, MS 100-22, Caltech, 1200 E.\ California Blvd., Pasadena, CA 91125, USA}

\author[0000-0001-7780-3352]{Michael C. Cushing}
\affiliation{University of Toledo, 2801 W. Bancroft Street, MS 113, Toledo, OH 43606, USA}

\author[0000-0002-7669-1069]{Sebastiano Calchi-Novati}
\affiliation{IPAC, MS 100-22, Caltech, 1200 E.\ California Blvd., Pasadena, CA 91125, USA}

\author{Y. Shvartzvald}
\affiliation{IPAC, MS 100-22, Caltech, 1200 E.\ California Blvd., Pasadena, CA 91125, USA}

\author[0000-0002-0638-8822]{Philip S. Muirhead}
\affiliation{Department of Astronomy \& Institute for Astrophysical Research, Boston University, 725 Commonwealth Ave., Boston, MA 02215, USA}




\begin{abstract}

M subdwarfs are low-metallicity M dwarfs that typically inhabit the halo population of the Galaxy. Metallicity controls the opacity of stellar atmospheres; in metal poor stars, hydrostatic equilibrium is reached at a smaller radius, leading to smaller radii for a given effective temperature. We compile a sample of 88 stars that span spectral classes K7 to M6 and include stars with metallicity classes from solar-metallicity dwarf stars to the lowest metallicity ultra-subdwarfs to test how metallicity changes the stellar radius. We fit models to Palomar Double Spectrograph (DBSP) optical spectra to derive effective temperatures ($T_\mathrm{eff}$) and we measure bolometric luminosities ($L_\mathrm{bol}$) by combining broad wavelength-coverage photometry with Gaia parallaxes. Radii are then computed by combining the $T_\mathrm{eff}$ and $L_\mathrm{bol}$ using the Stefan-Boltzman law. We find that for a given temperature, ultra-subdwarfs can be as much as five times smaller than their solar-metallicity counterparts. 
We present color-radius and color-surface brightness relations that extend down to [Fe/H] of $-$2.0 dex, in order to aid the radius determination of M subdwarfs, which will be especially important for the \textit{WFIRST} exoplanetary microlensing survey.

\end{abstract}


\section{Introduction}
\label{s:intro}

M subdwarfs are low-metallicity M-dwarf stars and are identified by their position to the left of the main sequence on a color magnitude diagram \citep{sandage59}. Their metal-poor compositions are a characteristic of their old age, and therefore M subdwarfs make up a significant portion of the halo and bulge stellar populations \citep[e.g., ][]{gizis97, lepine03, burgasser03}. The very low metallicity of the subdwarfs is theorized to alter their radii since metallicity controls the opacity of the atmosphere, which modifies the equilibrium configuration \citep{burrows93}. In metal-poor stars the photosphere is expected to lie deeper in the star where the gas temperature is higher, leading to smaller radii for a given effective temperature ($T_\mathrm{eff}$). 

Accurate stellar radii are extremely important for exoplanet characterization; improved radius measurements have enabled new discoveries of transiting exoplanets in the Kepler sample \citep[e.g., ][]{fulton17}. Although subdwarfs have not been targeted often by many transiting exoplanet surveys, their radii will be important for NASA's Wide Field Infrared Survey Telescope's ({\it WFIRST}) exoplanet microlensing survey. 
The survey is a wide-area microlensing study targeting source stars in the Galactic bulge. Bulge stars will be monitored via a wide near-infrared band for brightening indicative of lensing by an intervening foreground object. A planet in orbit around the lensing star can sometimes be detected as a secondary perturbation \citep{gaudi12}. By measuring many secondary events, {\it WFIRST} will perform a statistical census of the Galaxy’s planetary population in a way not possible with direct imaging or radial velocity techniques and in a way that samples a different parameter space than transit studies \citep{penny18}. 

Subdwarfs will represent a significant fraction of Galactic bulge sources observed during the exoplanetary microlensing survey. When these sources are brightened by foreground lensing systems containing one or more exoplanets, their accurate characterization is an important component in determining the properties of the lensing system itself. 
A large fraction of \textit{WFIRST}'s exoplanet microlensing events will display finite source effects \citep{zhu14}, where sharp features of the lens' magnification pattern resolve the finite angular size of the source star \citep[e.g., ][]{witt94} and allow measurement of the ratio of the angular source radius to the angular Einstein radius. Knowledge of the angular source radius, e.g., from use of color-surface brightness relations \citep{yoo04, kervella08, boyajian12} allows the ratio to be converted into a measurement of the angular Einstein radius and a constraint on the mass of the lens \citep{gould94, nemiroff94}.
Yet, the sizes of subdwarfs are not well known, mainly because subdwarfs are rare in the solar neighborhood and have not seen the scrutiny that stars of higher metallicity have seen.

Previous studies have discovered and classified many M subdwarfs, but less has been done to determine their physical parameters (e.g., radii and effective temperatures). \citet{gizis97} first introduced a classification scheme for M subdwarfs based on the molecular line strength ratios between the optical CaH ($\sim$6830 and 6975 \AA) and TiO5 ($\sim$7130 \AA) bands and separated M subdwarfs into three categories: the solar metallicity dwarfs (dM), the metal-poor subdwarfs (sdM), and the very metal-poor extreme subdwarfs (esdM). \citet{lepine07} increased the sample of known metal poor M dwarfs to over 400 objects and expanded the classification to include a new even more metal-poor class, ultra subdwarfs (usdM). 

Since then, \citet{jao08} devised a separate classification scheme for subdwarfs, based on physical parameters (effective temperature, metallicity and surface gravity), by comparing spectra to stellar atmosphere models.
Exact values of these physical parameters could not be determined until recently because model atmospheres still could not reproduce many of the molecular features present in the atmospheres of cool stars. However, \citet{rajpurohit14, rajpurohit16} found that the recently updated PHOENIX stellar atmosphere models \citep{allard12} successfully reproduced many of the features in low metallicity stars and were therefore able to make estimates of the metallicity, surface gravity and temperature of a limited sample of M subdwarfs. 

Recently, there has also been a significant effort to expand the sample of subdwarfs to the very lowest mass stars and brown dwarfs \citep[e.g., ][]{zhang17, zhang18}. \citet{zhang18} increased the known sample of L subdwarfs to about 66 objects that have been spectroscopically confirmed and classified. \citet{gonzales18} determined fundamental parameters (e.g., temperature, bolometric luminosity) for 10 of these L subdwarfs. These studies are complementary to our work since they focus on stars of spectral type M7 through L, while our targets are spectral type K7 through M7. Together, a temperature sequence from K7 through L-type metal-poor stars and brown dwarfs can be created.

In this paper we present stellar radii for a greatly expanded sample of M subdwarf stars. In Section \ref{s:Sample} we describe how we chose our representative sample of M subdwarf stars, and in Section \ref{s:Data} we describe our Palomar DBSP observations and data reduction procedure. The radii are calculated by combining $T_\mathrm{eff}$ and $L_\mathrm{bol}$ using the Stefan-Boltzmann equation. We detail our method for determining the metallicity in Section \ref{s:Metal}, the effective temperature in Section \ref{s:Teff}, and our method for determining the bolometric luminosity in Section \ref{s:Lbol}. Finally, we present color and effective temperature relations that can be used to determine the radii of other M subdwarf stars in Section \ref{s:Results}.

\section{Selecting the Sample}
\label{s:Sample}

The {\it WFIRST} microlensing survey will probe sources primarily in the 20 $<$ W149\footnote{This is a wide filter extending from 0.927 to 2.000 $\mu$m. See the list of {\it WFIRST} telescope and instrument parameters at \url{http://wfirst.ipac.caltech.edu}} $<$ 24 mag (AB) range, corresponding roughly to early-G through mid-M spectral types at the 8 kpc distance of the Galactic bulge, assuming a total column extinction of A(W149) $\approx$ 1.0 mag toward $l=1{\fdg}0$, $b=-1{\fdg}5$ (\citealt{schlafly11}). The metallicity range of stars in the Galactic bulge spans $-3.0 < [Fe/H] < 1.0$ dex (\citealt{ness16}). The more metal-rich stars in this range are those that trace out the well-known boxy/peanut shape of the inner Galaxy. The more metal-poor stars belong either to a thick disk or an old spheroidal population (\citealt{dekany13, gran16}). Therefore, we wish to use observations of bright, nearby subdwarfs to construct a grid of spectra covering the spectral type and metallicity range present in the bulge that, when combined with photometry at other wavelengths, will allow us to fully characterize a broad subset of these objects. Knowledge learned from this nearby subset can then be used to deduce radii for more distant examples using color information alone.

Subdwarfs first become identifiable in broadband colors at mid-K types (see Figure 1 from \citealt{zhang17} and Figure 7 from \citealt{kirkpatrick16}). The proposed \textit{WFIRST} microlensing observations will probe bulge dwarfs as cold as roughly mid-M. Therefore, we restricted our spectral class range to $\sim$K7 through $\sim$M7. The \cite{lepine07} subdwarf subclasses – subdwarfs (sd), extreme subdwarfs (esd), and ultra subdwarfs (usd), roughly represent objects in the metallicity ranges log($[Fe/H]$) $\approx$ $-$0.5, $-$1.0, and $-$1.5, respectively. Most known late-K through late-M subdwarfs were classified before the Lepine et al. system was established, some of which were typed against the earlier \cite{gizis97} two-subclass system. Others pre-date both of these papers and are classified on a mixture of systems. 

\begin{figure}[h]
\begin{center}
\includegraphics[width=\linewidth]{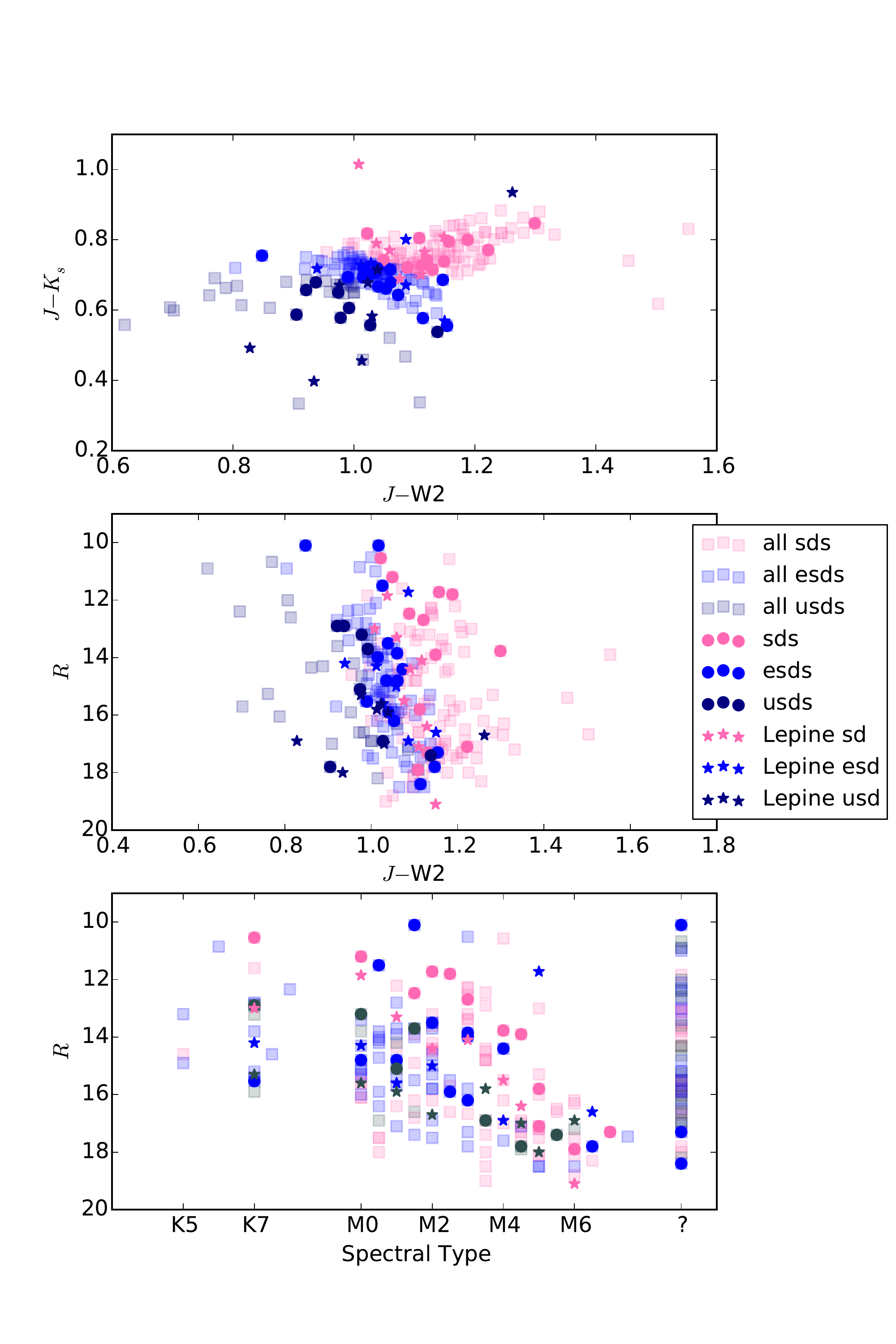}
\caption{\small 
\textbf{Top:} $J-K_s$ versus $J-$W2 diagram, used to separate the compiled $\sim$250 selected subdwarfs into the metallicity classes of d, sd, esd, and usd. The targets ultimately selected are colored circles, the \cite{lepine07} subdwarf standards are shown as colored stars, and the full original sample is shown as translucent squares. Note that one of the Lepine usd standards has dwarf-like colors; this star is LHS 1691 and we believe that its 2MASS $J$-band color is not correct. This star is also an outlier in later figures, such as Figure \ref{f:ColorR}. \textbf{Middle:} $R$ magnitude versus $J-$W2 color diagram. \textbf{Bottom:} $R$-band magnitude versus spectral type diagram. A target without a known spectral type is shows as a `?' on this plot. This plot illustrates how we tried to target two bright sds, esds, and usds for each spectral type estimate. }
\label{f:sampleSelect}
\end{center}
\end{figure}

Rather than rely on published types, we combed the literature for objects classified as subdwarfs. We identified $\sim$250 in all, most of which are relatively bright, nearby sources found by various proper motion surveys. We then tabulated their optical, 2MASS, and {\it WISE} magnitudes.  Using the $J-K_s$ vs. $J-$W2 diagram, we plotted these objects together with known dwarfs of solar metallicity, the subdwarf standards of \cite{lepine07}, and the theoretical subdwarf tracks (see Figure 1 from \citealt{zhang17}) to pseudo-categorize each as d, sd, esd or usd. This color-color diagram is shown in the top plot of Figure \ref{f:sampleSelect}. After removing those that appeared to be solar-metallicity dwarfs and those too far south to be observed with the 200 inch telescope at Palomar, we were able to sort the distribution of candidates by $R$ magnitude and $J-$W2 color, the latter being a proxy for temperature or spectral type. Using this list, we created a target list having three objects in each integral spectral type bin. Three objects per bin were required to mitigate the effects of unresolved binarity on the $L_{bol}$ determination and to have a crude assessment of the cosmic scatter per bin. One object in each bin was chosen to be the \cite{lepine07} standard itself, and the other two were generally chosen to be the brightest (and therefore most easily observable) at $R$ band. This final observing list is shown in Table~\ref{target_grid} as well as in Figure \ref{f:sampleSelect}.

Prior to our spectroscopic observations, we created finder charts at the 2017.8 epoch of each source, using the source's 2MASS position and its published proper motion. Any source confused with a bright background source at our epoch was replaced with the next brightest star in the spectral bin. One of the subdwarf standards, LSR J1918+1728 (esdM3), is contaminated at our epoch of observation and was therefore skipped.

In order to facilitate spectral classification comparisons and to provide checks of radius measurements for stars similar to those in \cite{mann13}, we observed two to three solar metallicity dwarfs in each spectral subtype bin, as well. These are also listed in Table~\ref{target_grid}.

\begin{center}
\begin{deluxetable*}{lllll}
\tabletypesize{\small}
\tablecaption{Spectral Type Grid\label{target_grid}}
\tablehead{
\colhead{Sp.\ Type} &                          
\colhead{Dwarfs} &
\colhead{Subdwarfs} &     
\colhead{Extreme}  &
\colhead{Ultra} \\
\colhead{Range} &                          
\colhead{} &
\colhead{} &     
\colhead{Subdwarfs}  &
\colhead{Subdwarfs} \\ 
\colhead{(1)} &                          
\colhead{(2)} &  
\colhead{(3)} &     
\colhead{(4)} &
\colhead{(5)}    
}
\startdata
K7-8   &  Gl 143.1  &  LHS 1703*         & LHS 3276*         &  LHS 1454*       \\
       &  ---  &  LHS 170           & LHS 104           &  LSR J0621+3652  \\
       &  ---   &  LHS 173           & LHS 522           &  LSR J2115+3804  \\   
M0-0.5 &  Gl 270*   &  LHS 12*           & LHS 360*          &  LHS 2843*       \\   
       &  ---   &  LHS 42            & LHS 489           &  LHS 182         \\
       &  ---       &  LHS 174           & LHS 2355          &  LSR J1956+4428  \\    
M1-1.5 &  Gl 229A*  &  LHS 2163*         & LHS 1994*         &  LHS 1863*       \\ 
       &  Gl 908    &  LHS 482           & LHS 364           &  LHS 518         \\   
       &  ---       &  LHS 178           & LHS 318           &  LSR J2205+5353  \\   
M2-2.5 &  Gl 411*   &  LHS 228*          & LHS 2326*         &  LHS 1691*       \\ 
       &  Gl 393    &  LHS 2852          & LHS 3555          &  LSR J0020+5526  \\   
       &  ---       &  LHS 20            & LHS 161           &  WISE J0707+1705 \\
M3-3.5 &  Gl 436*   &  LSR J0705+0506*   & [LSR J1918+1728*] &  [LHS 325*]      \\ 
       &  Gl 109    &  LHS 272           & LHS 1174          &  LSR J0522+3814  \\
       &  Gl 388    &  LHS 156           & LHS 3263          &  LHS 3382        \\   
M4-4.5 &  Gl 402*   &  LHS 2674*/LHS 504*& LSR J1340+1902*   &  LHS 1032*       \\   
       &  Gl 447    &  NLTT 3247         & LHS 375           &  LHS 4028        \\       
       &  LHS 3255  &  LHS 3409          & LHS 3090          &  LHS 453         \\   
M5-5.5 &  Gl 51*    &  LHS 2061*         & LHS 2405*         &  LHS 2500*       \\ 
       &  [LP 467-16] &  LHS 3189          & LHS 515           &  LSR J2122+3656  \\   
       &  ---       &  LHS 3390          & LHS 2096          &  LHS 205a        \\
M6-6.5 &  Gl 406*   &  [LHS 2746*]       & LHS 2023*         &  LSR J0621+1219* \\ 
       &  Teegarden &  LHS 1166          & 2MASS J0822+1700  &  LHS 1625        \\
       &  ---       &  LHS 1074          & LHS 1742a         &  LHS 1826        \\    
M7-7.5 &  ---       &  LHS 377           & ---               &  ---             \\
\enddata
\tablecomments{An asterisk indicates a spectral standard. The three spectral standards in braces were not, however, observed: LSR J1918+1728 because it was confused at our observation epoch with a background star, LHS 2746 because it was too faint for the observing conditions, and LHS 325 because of a typographical error in our observing list. LP 467-16 was observed but was later determined to be a binary and we therefore do not list parameters for it. A few of the object names are abbreviated in the table: ``Teegarden'' is Teegarden's Star; ``2MASS J0822+1700'' is 2MASS J08223369+1700199, and ``WISE J0707+1705'' is WISEA J070720.50+170532.7.}
\end{deluxetable*}
\end{center}

\section{Observations and Data Reduction}
\label{s:Data}

Data were taken during six separate nights between August of 2017 and January of 2018, using DBSP on the 200-inch Hale Telescope at Palomar Observatory. DBSP is a moderate resolution optical spectrograph that uses a dichroic to split light into separate red and blue channels that are observed simultaneously \citep{oke82}. The observer can choose from four different dichroics and can choose the grating angle to set the wavelength coverage and spectral resolution. For all of our nights we chose the dichroic that split the light at a wavelength of 6800\AA. For the blue side we used a 600/4000 grating and for the red side a 600/10000 grating. We chose grating angles of $\sim29^{\circ}$ and $\sim32^{\circ}$, leading to a wavelength coverage of $\sim3900-6950$\AA\ and $\sim6610-9970$\AA\, and a mean resolving power of $\sim$2,000 and $\sim$3,000 for the blue and red sides, respectively. 

We performed all of the data reduction using the python command line tool for IRAF (PyRAF). Bias subtraction, flat fielding, spectral extraction, cosmic ray removal, wavelength calibration and flux calibration were performed on the red and blue images separately. Wavelength calibration frames using a Fe-Ar lamp for the blue side and a He-Ne-Ar lamp for the red side were taken at the beginning of each night. 

\begin{figure*}
\plotone{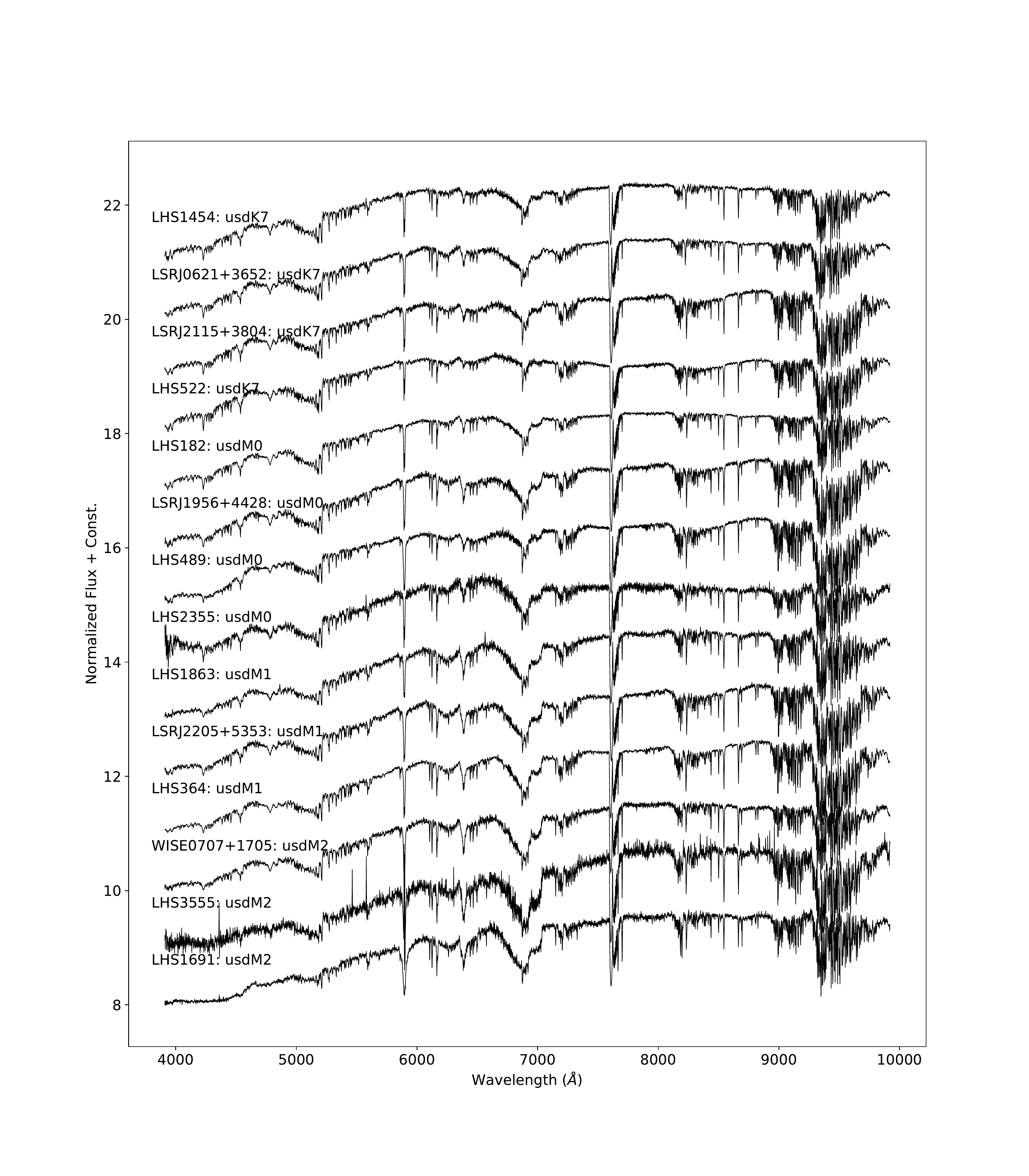}
\caption{Reduced, flux-calibrated spectra of the early-type (K7-M2) ultra subdwarfs in our sample. All the spectra are available in the online article using the data behind the figures feature.}
\label{f:spectra}
\end{figure*}

The red and blue wavelength scales were each shifted to rest separately by cross correlation with a model stellar spectrum of spectral type either M1 for the hotter stars, or M5 for the cooler stars. We next stitched the spectra together by normalizing the spectra to each other at the stitch point. The stitch point was chosen by visual inspection of each spectrum to be a point with relatively low noise and free of any large absorption features, and fell between  $6650-6775$\AA. All the spectra are available in Figure Set \ref{f:spectra} (8 images), and available in the online journal. 

\figsetstart
\figsetnum{2}
\figsettitle{Subdwarf Spectra}

\figsetgrpstart
\figsetgrpnum{2.1}
\figsetgrptitle{Early-type usd Spectra}
\figsetplot{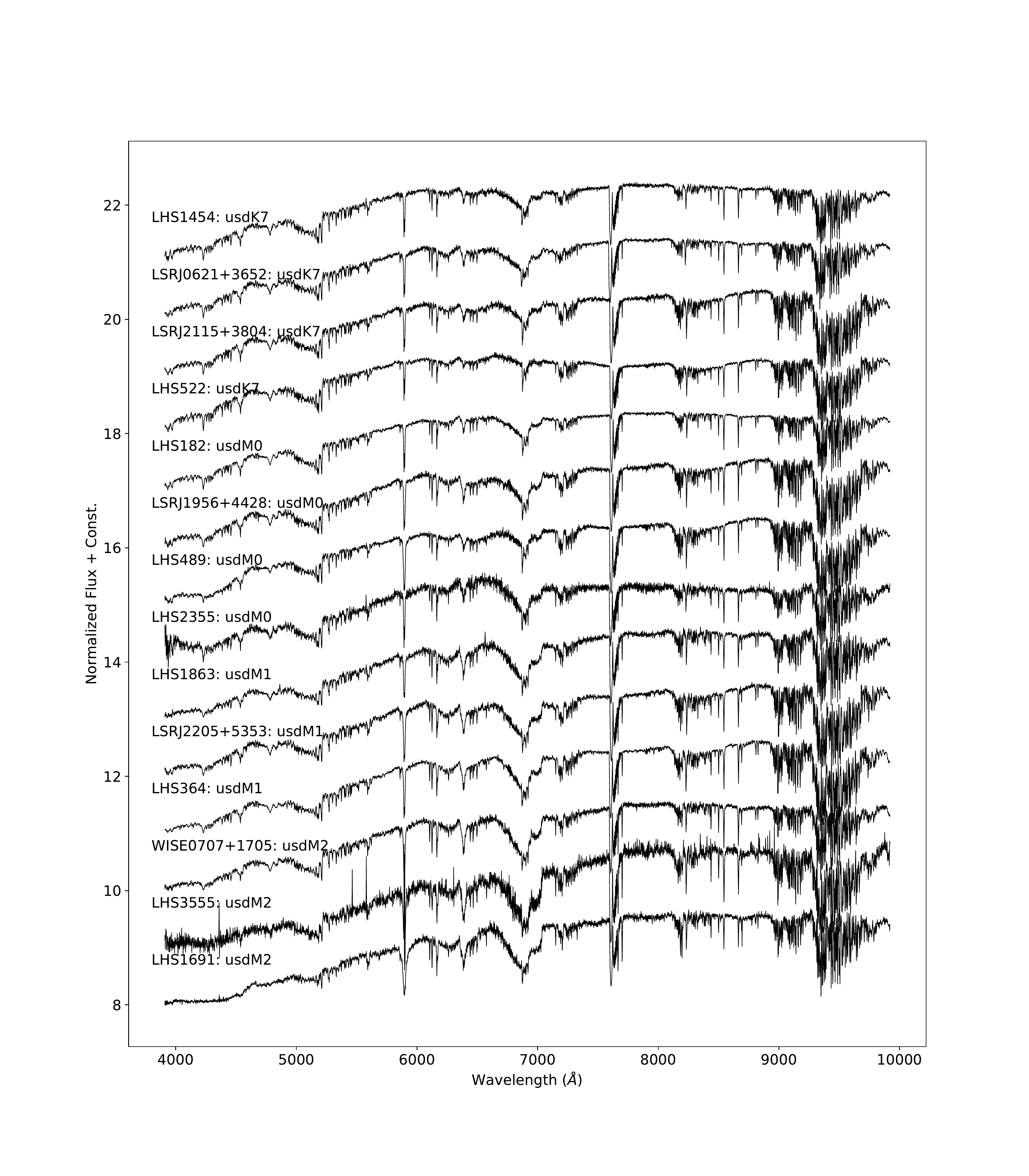}
\figsetgrpnote{Reduced, flux-calibrated spectra of the early-type (K7-M2) ultra subdwarfs in our sample. All the spectra are available in the online article using the data behind the figures feature.}
\figsetgrpend

\figsetgrpstart
\figsetgrpnum{2.2}
\figsetgrptitle{Late-type usd Spectra}
\figsetplot{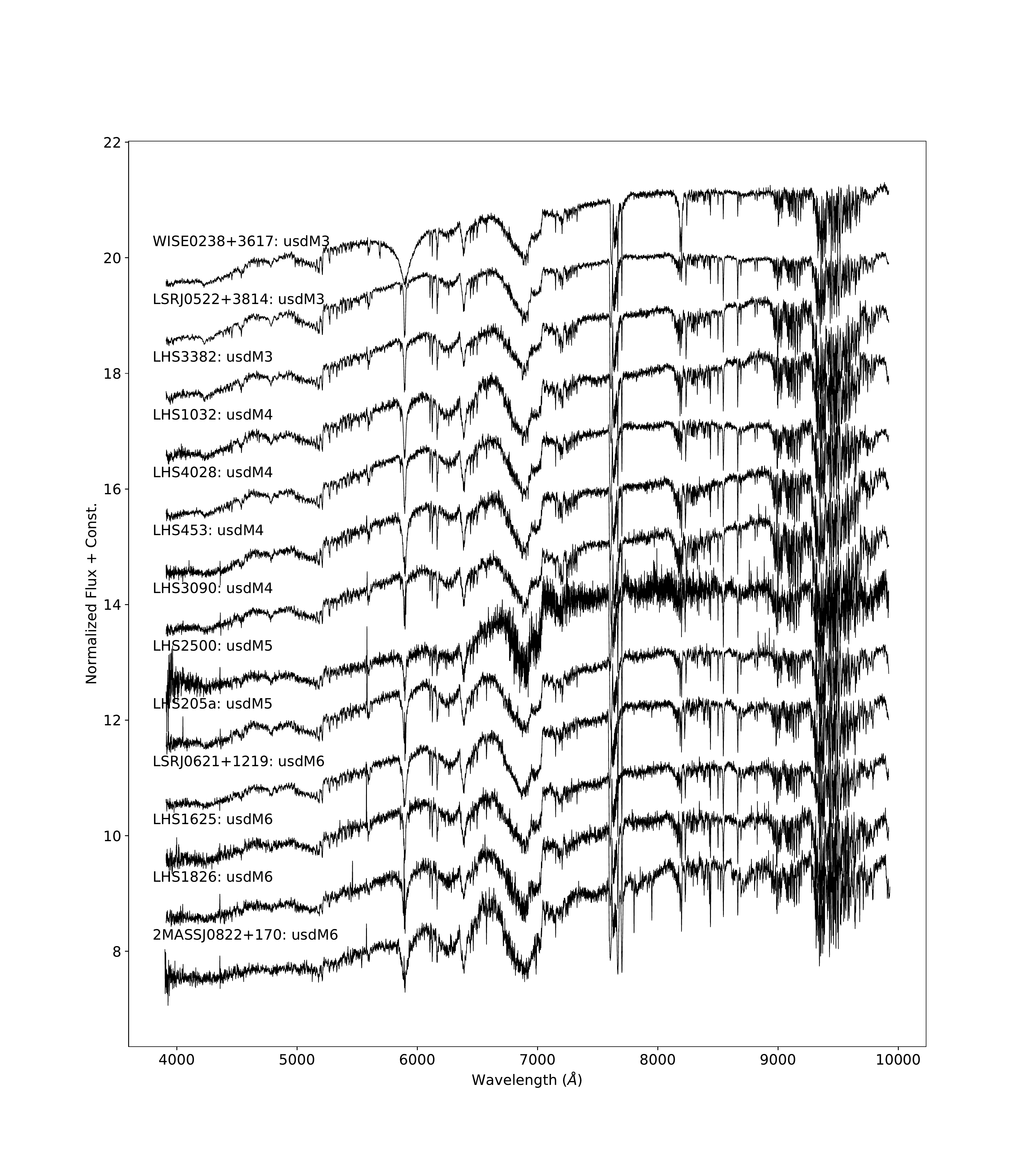}
\figsetgrpnote{Reduced, flux-calibrated spectra of the late-type (M3-M6) ultra subdwarfs in our sample. All the spectra are available in the online article using the data behind the figures feature.}
\figsetgrpend

\figsetgrpstart
\figsetgrpnum{2.3}
\figsetgrptitle{Early-type esd Spectra}
\figsetplot{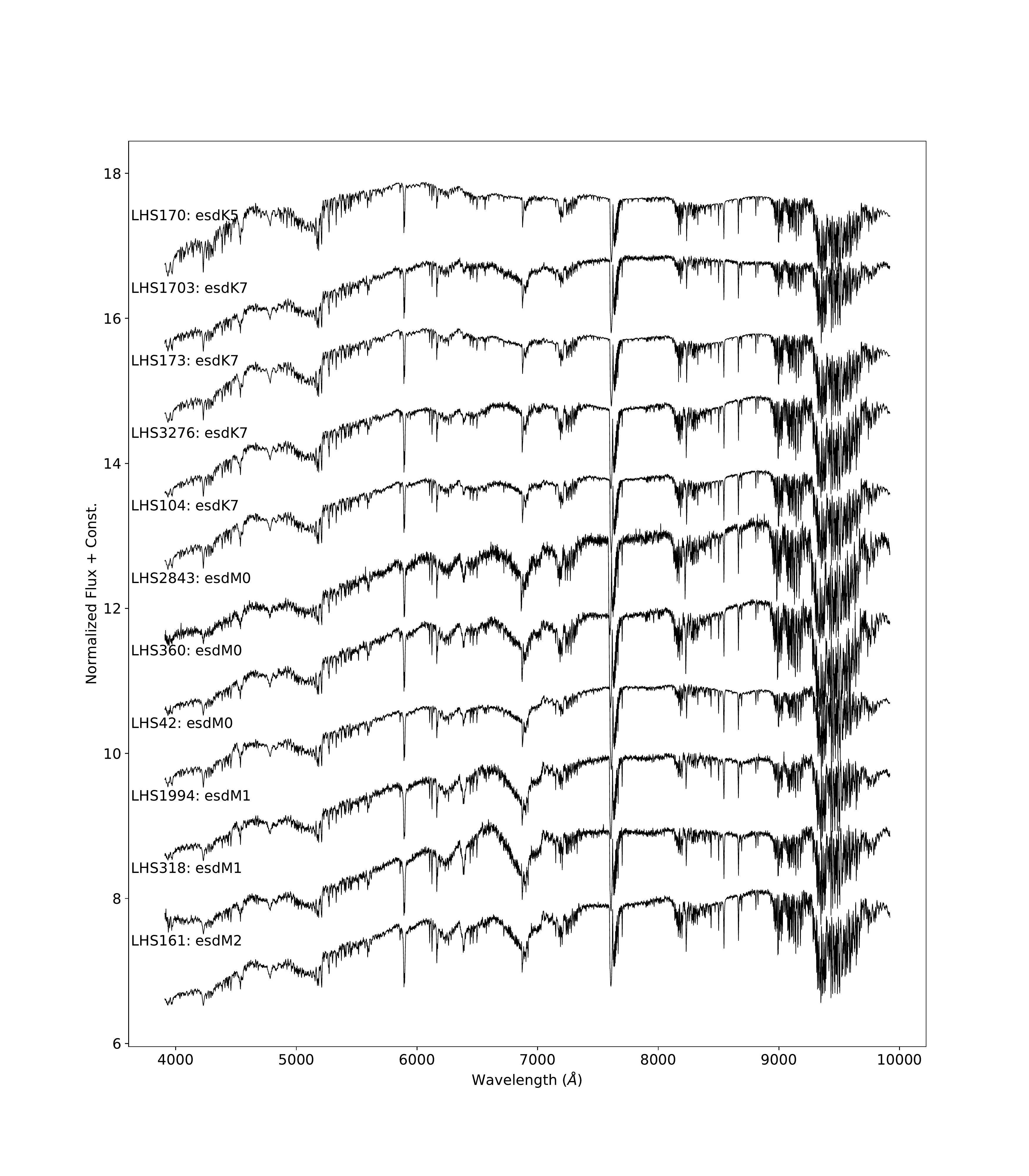}
\figsetgrpnote{
Reduced, flux-calibrated spectra of the early-type (K7-M3) extreme subdwarfs in our sample. All the spectra are available in the online article using the data behind the figures feature.}
\figsetgrpend

\figsetgrpstart
\figsetgrpnum{2.4}
\figsetgrptitle{Late-type esd Spectra}
\figsetplot{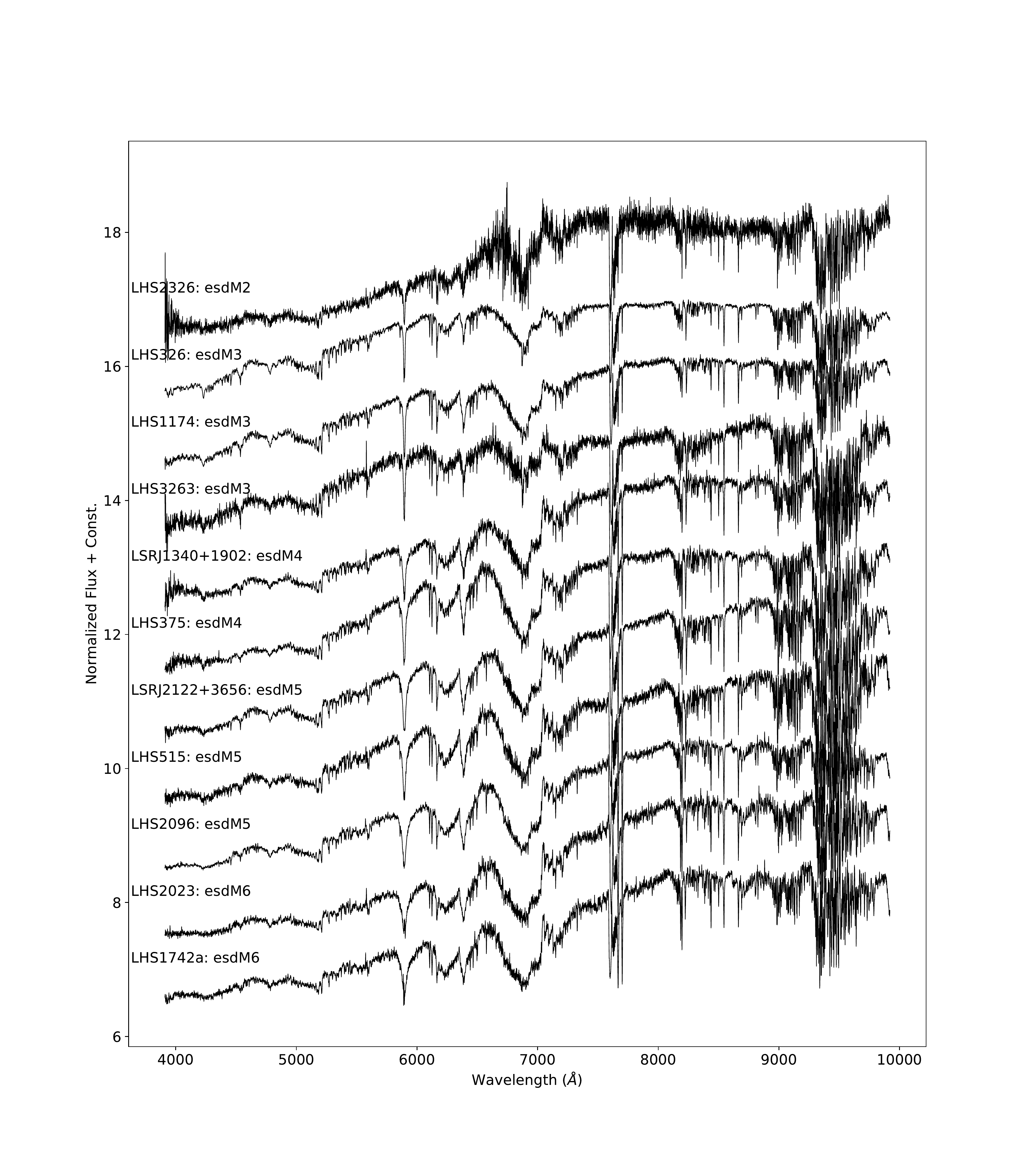}
\figsetgrpnote{
Reduced, flux-calibrated spectra of the late-type (M3-M6) extreme subdwarfs in our sample. All the spectra are available in the online article using the data behind the figures feature.}
\figsetgrpend

\figsetgrpstart
\figsetgrpnum{2.5}
\figsetgrptitle{Early-type sd Spectra}
\figsetplot{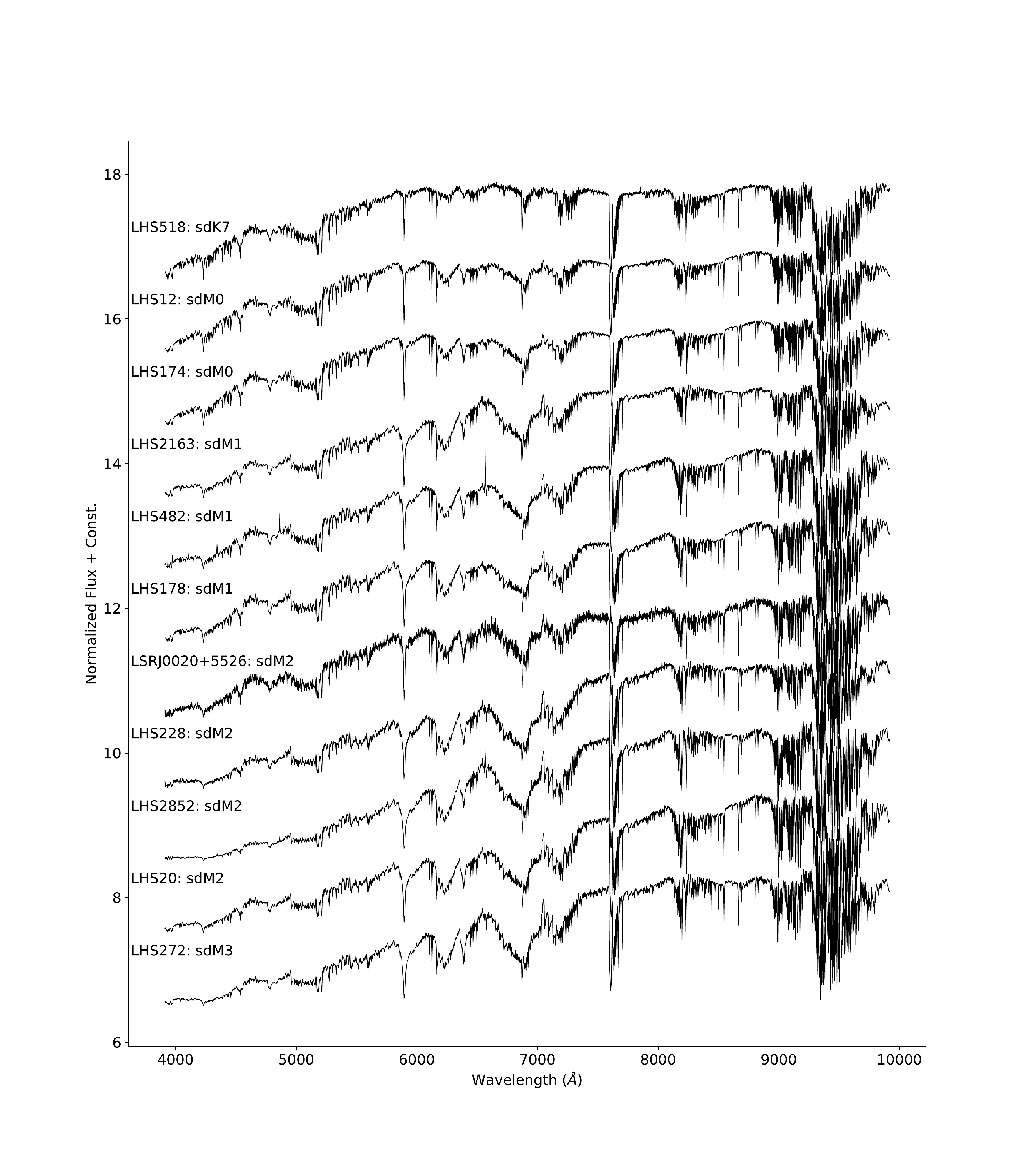}
\figsetgrpnote{
Reduced, flux-calibrated spectra of the early-type (K7-M2) subdwarfs in our sample. All the spectra are available in the online article using the data behind the figures feature.}
\figsetgrpend

\figsetgrpstart
\figsetgrpnum{2.6}
\figsetgrptitle{Late-type sd Spectra}
\figsetplot{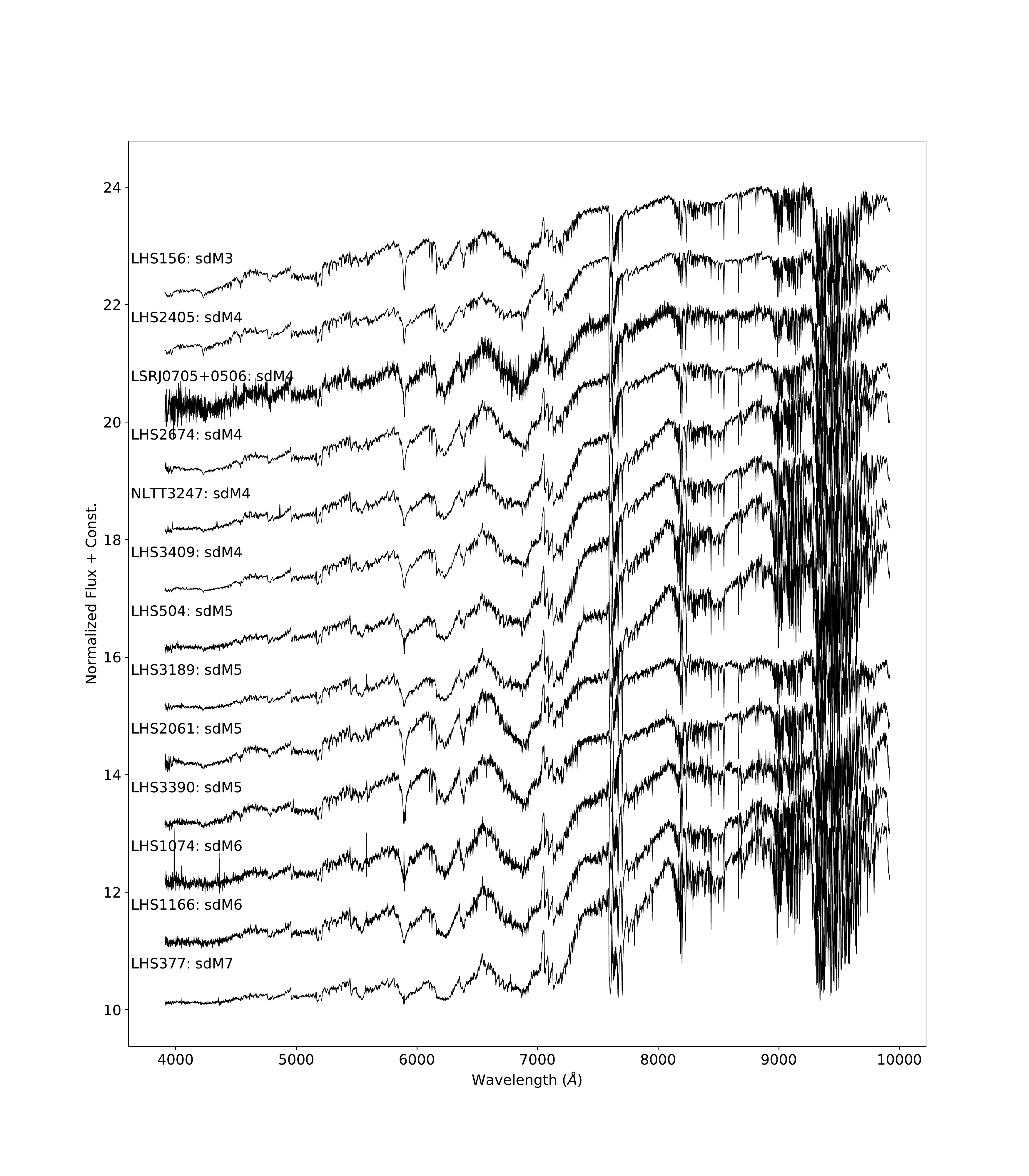}
\figsetgrpnote{
Reduced, flux-calibrated spectra of the late-type (M3-M7) subdwarfs in our sample. All the spectra are available in the online article using the data behind the figures feature.}
\figsetgrpend

\figsetgrpstart
\figsetgrpnum{2.7}
\figsetgrptitle{Early-type d Spectra}
\figsetplot{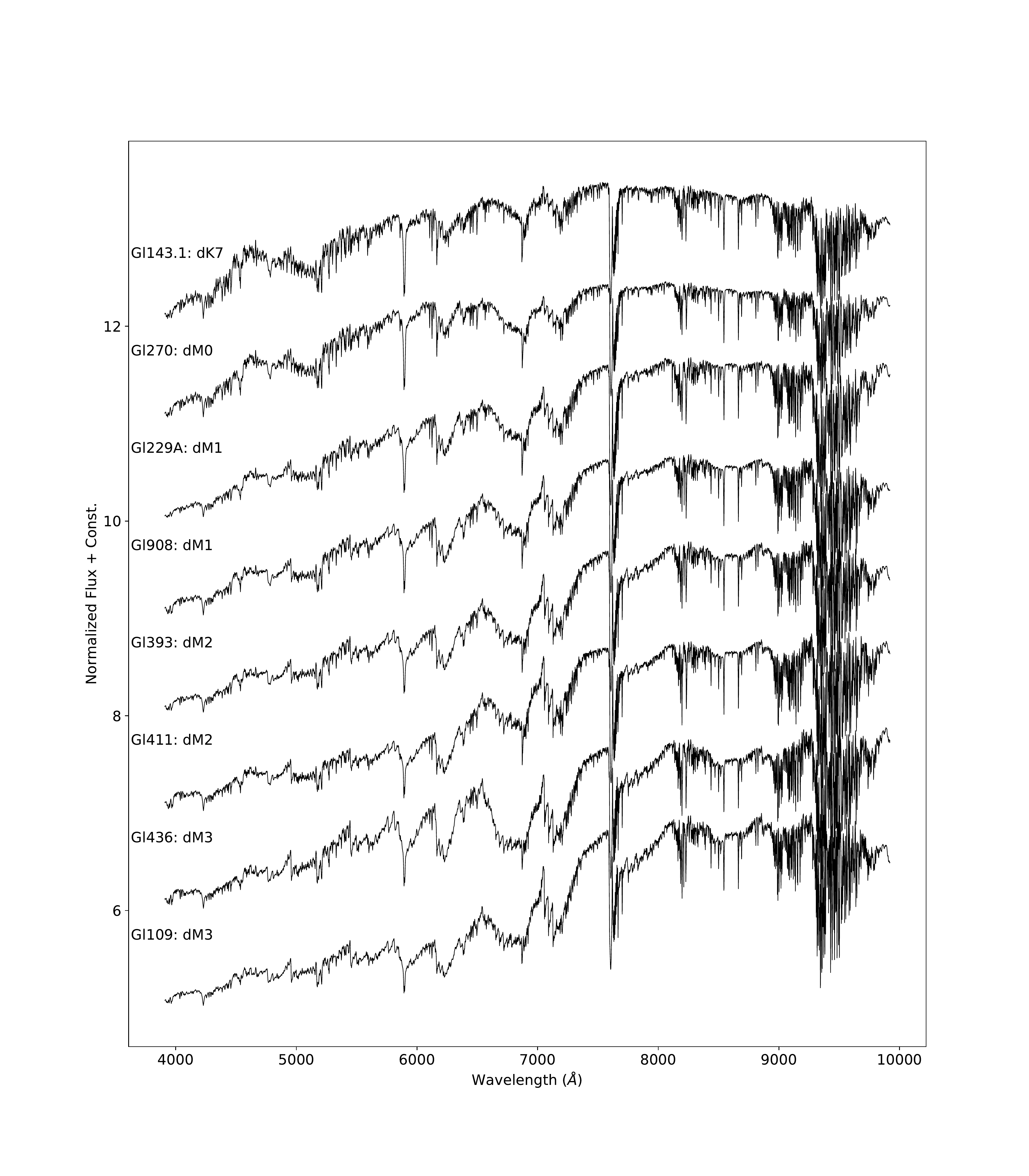}
\figsetgrpnote{
Reduced, flux-calibrated spectra of the early-type dwarfs (K7-M3) in our sample. All the spectra are available in the online article using the data behind the figures feature.}
\figsetgrpend

\figsetgrpstart
\figsetgrpnum{2.8}
\figsetgrptitle{Late-type d Spectra}
\figsetplot{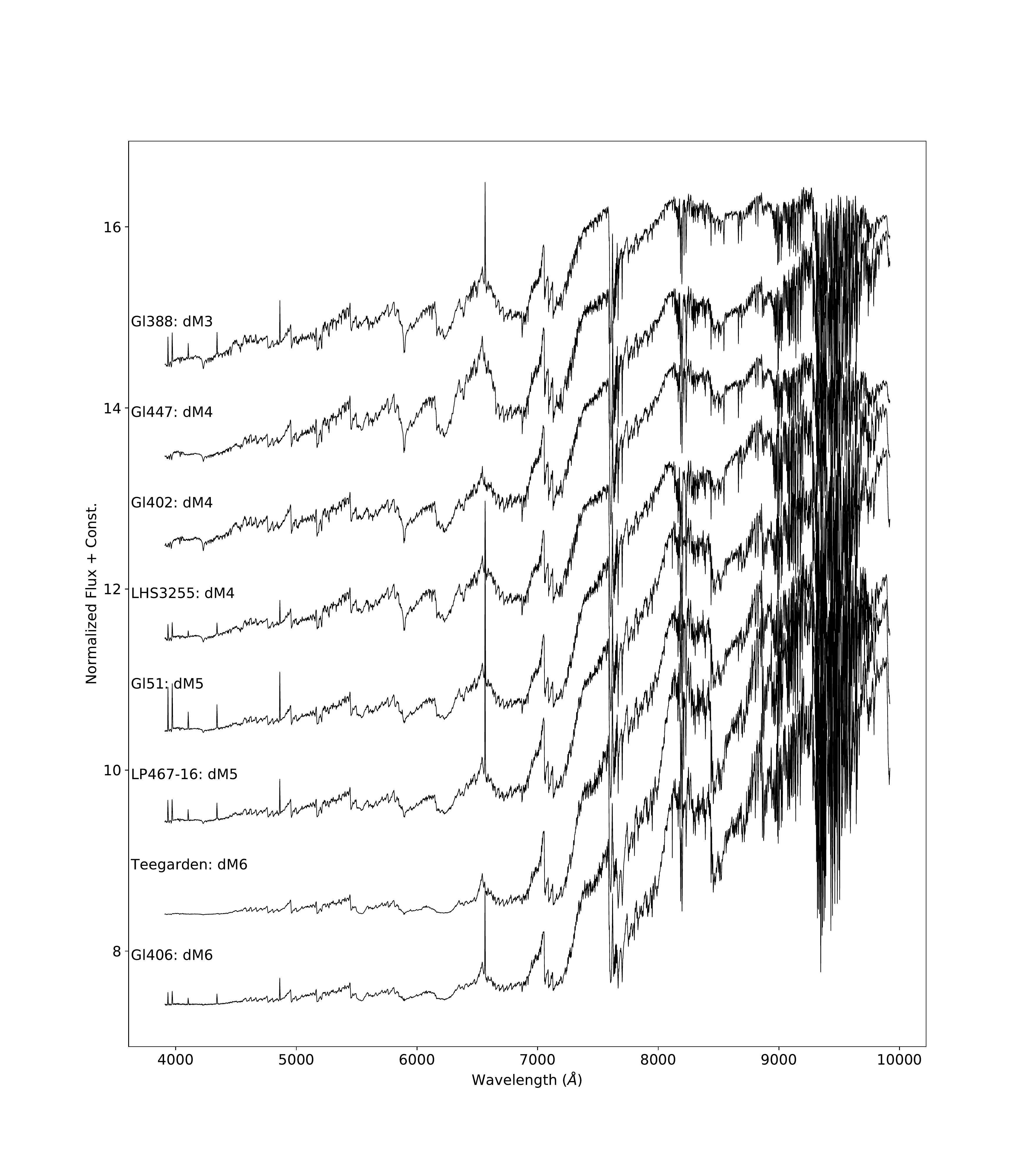}
\figsetgrpnote{
Reduced, flux-calibrated spectra of the late-type dwarfs (M3-M6) in our sample. All the spectra are available in the online article using the data behind the figures feature.}
\figsetgrpend

\figsetend

For a small subset of our targets, we also obtained high resolution near-infrared spectra from iSHELL \citep{rayner12} on NASA's 3.0-meter Infrared Telescope Facility (IRTF) on Mauna Kea, Hawaii. We used the wider slit width, giving a spectral resolution of about R$\sim$35,000 for our chosen wavelength region ($2.09-2.38 \mu m$). In total we collected spectra of three dwarfs, four subdwarfs, one extreme subdwarf and one ultra subdwarf, to test our metallicity estimate techniques (see Section \ref{s:Metal} for details). We completed the data reduction of the iSHELL spectra using the Spextool for iSHELL package\footnote{\url{http://irtfweb.ifa.hawaii.edu/research/dr_resources/}}. Spextool \citep{cushing04} has been updated in the newest release to be compatible with iSHELL data, and performs dark subtraction, flat fielding, order tracing and extraction, linearity correction and returns a wavelength solution calibrated using ThAr lamps. We removed telluric absorption features using the \texttt{xtellcor} \citep{vacca03} function, which is also part of the larger Spextool reduction package.

\section{Determining Stellar Parameters}

\subsection{Metallicity}
\label{s:Metal}

Precise metallicities of M dwarfs are notoriously difficult to determine because much of the spectrum is dominated by deep molecular features resulting in a lack of a true continuum in much of the spectrum. Recently however, many groups have successfully used widely separated binaries or common proper motion stars that contain an F, G, or K star and an M dwarf companion to calibrate methods that use metallicity sensitive spectral features to determine precise metallicities of M dwarfs \citep[e.g., ][]{rojas10, rojas12, terrien12, mann13, newton14}. Unfortunately, all of the relations presented in these studies focus on solar-metallicity or near-solar metallicity stars and are not calibrated for the low metallicities present in our sample. We therefore use two different methods: one to determine the metallicity of the dwarf and dwarf/subdwarf stars ([Fe/H] $>-0.5$ dex), and a second to determine the metallicities of the more metal poor subdwarfs, and the extreme and ultra subdwarfs. 

\begin{figure}[h]
\begin{center}
\includegraphics[width=\linewidth]{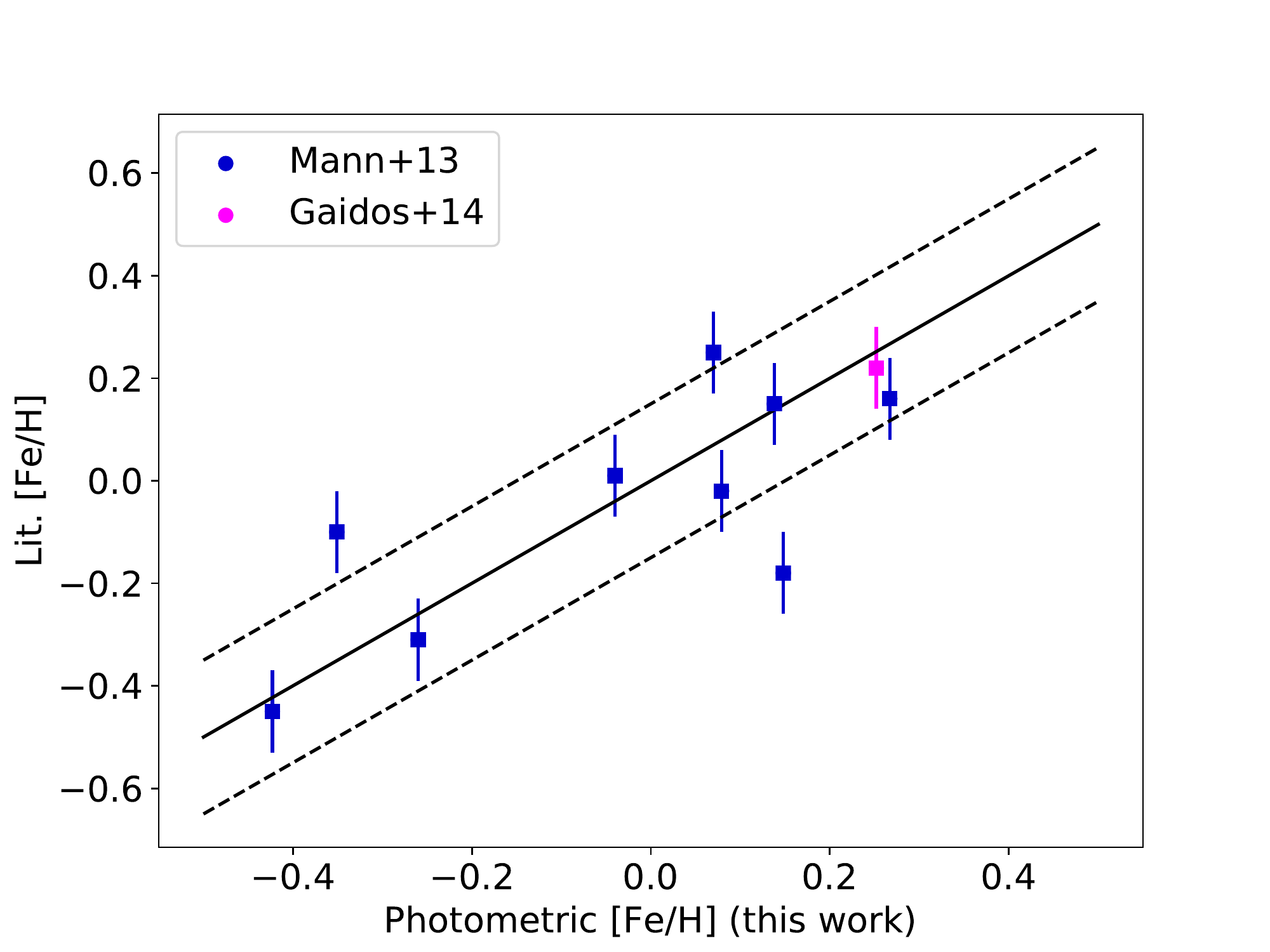}
\caption{\small 
Comparison between our values of [Fe/H] and previously measured literature [Fe/H] values for 10 of our objects. Our values of [Fe/H] were determined photometrically using the near-IR color to metallicity relation from \citet{newton14}. The literature values of [Fe/H] were determined spectroscopically by \citet{gaidos14} and \citet{mann15}, both using the method outlined in \citet{mann13}. We find that the photometric metallicities show the same trend as the spectroscopic metallicities and that there is no bias towards over or underestimating the metallicities using photometric relations. The black solid line represents a one-to-one fit, and shows where all the points would lie if our photometrically determined [Fe/H] values matched the literature values exactly. We find a mean scatter around this line of 0.15 dex, and we adopt this value as our uncertainty for all of our values of [Fe/H] determined using this method.  }
\label{f:litMetal}
\end{center}
\end{figure}

The majority of the previously-mentioned methods use spectral features in the near-infrared, while our spectra are optical. \citet{mann13} published [Fe/H] relations that utilize optical spectra; however, the relations are highly dependent on the Na doublet at 8200\AA, which is contaminated by telluric features in our spectra and therefore it is difficult to measure an equivalent width. Because of this, we use the the near-infrared color relation from \citet{newton14} to estimate [Fe/H] for all the dwarfs and subdwarfs in our sample. Figure \ref{f:litMetal} shows how the photometric [Fe/H] compares to spectroscopic estimates of [Fe/H] from \citet{gaidos14} and \citet{mann15} for our 10 overlapping objects. We find a mean scatter of 0.15 dex and we adopt this as the uncertainty in [Fe/H] for the dwarfs and subdwarfs. 

Low-metallicity extreme and ultra subdwarfs are often categorized using a $\zeta$ parameter, which relates the CaH2 (6814$-$6846 \AA) and CaH3 (6960$-$6990 \AA) band ratios to the TiO5 (7126$-$7135 \AA) band, since the CaH band is primarily sensitive to temperature while the TiO5 band is sensitive to both temperature and metallicity \citep{dhital12}. Using high resolution spectra of subdwarfs and extreme subdwarfs, \citet{woolf09} determined a relationship between $\zeta$ and [Fe/H]. We made use of this relation and measured a $\zeta$ value and hence [Fe/H] for each of the stars in our sample. The relation was recalibrated by \citet{mann13}, but we find that the change in the derived value of [Fe/H] is significantly smaller than the quoted uncertainty of the relation (0.3 dex), and so we report the original [Fe/H] values determined with the \citet{woolf09} relation.

\begin{figure}[h]
\begin{center}
\includegraphics[width=\linewidth]{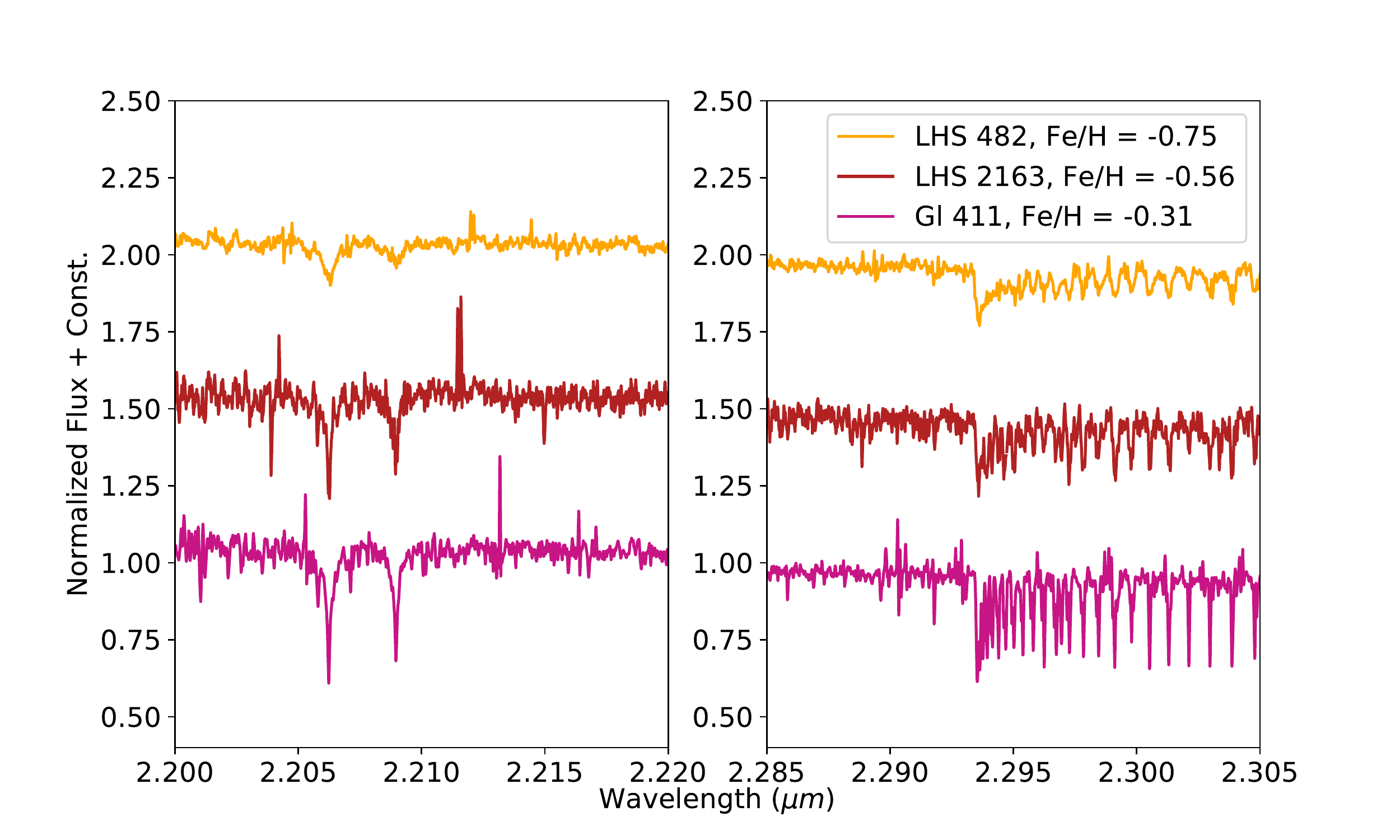}
\caption{\small 
 iSHELL $K$-band spectra of Gl 411, LHS 2163, and LHS 482. The spectra have all been shifted to rest by cross correlation with model templates. The left plot is centered on the sodium doublet (2.2 $\mu m$) and the right plot is centered on the CO bandhead (2.3 $\mu m$). These plots show the effect of decreased metallicity on these line strengths and how we can use the sodium doublet to estimate the stellar metallicity. We also note that LHS 482 seems to be rotationally broadened, which is intriguing since low metallicity ($-0.75$ dex) is reminiscent of old age while rapid rotation is reminiscent of youth \citep{west15}. This is the only star in our iSHELL sample which shows rotational broadening and we merely note it here as a potential future target of interest. }
\label{f:ishell}
\end{center}
\end{figure}

As an extra check, we used the high-resolution (R$\sim$35,000) near infrared iSHELL spectra of three dwarfs, four subdwarfs, one extreme subdwarf and one ultra subdwarf, to test the metallicities determined with the above methods.
Figure \ref{f:ishell} shows an example of our high resolution spectra and how the sodium doublet changes with metallicity. We calculated metallicities using the relation presented in \citet{newton14} that uses the equivalent width of the sodium doublet at 2.2 $\mu m$ to determine the metallicity with an uncertainty of 0.12 dex. We find that these metallicities agreed with the metallicities previously reported by \citet{mann13} for the three dwarf stars, and that the metallicities that we derive from the high resolution spectra are consistent with the metallicities derived using the \citet{woolf09} relation. One of our extreme subdwarfs (LHS 173) has a metallicity reported from the APOGEE Stellar Parameters and Chemical Abundances Pipeline (ASPCAP) \citep{schmidt16}. Our derived metallicity from the $\zeta$ parameter and the metallicity from (ASPCAP) are within 0.05 dex, which further validates our derived metallicites. 

\subsection{Effective Temperatures}
\label{s:Teff}

\begin{figure*}[htp]
\begin{center}
\includegraphics[width=\linewidth]{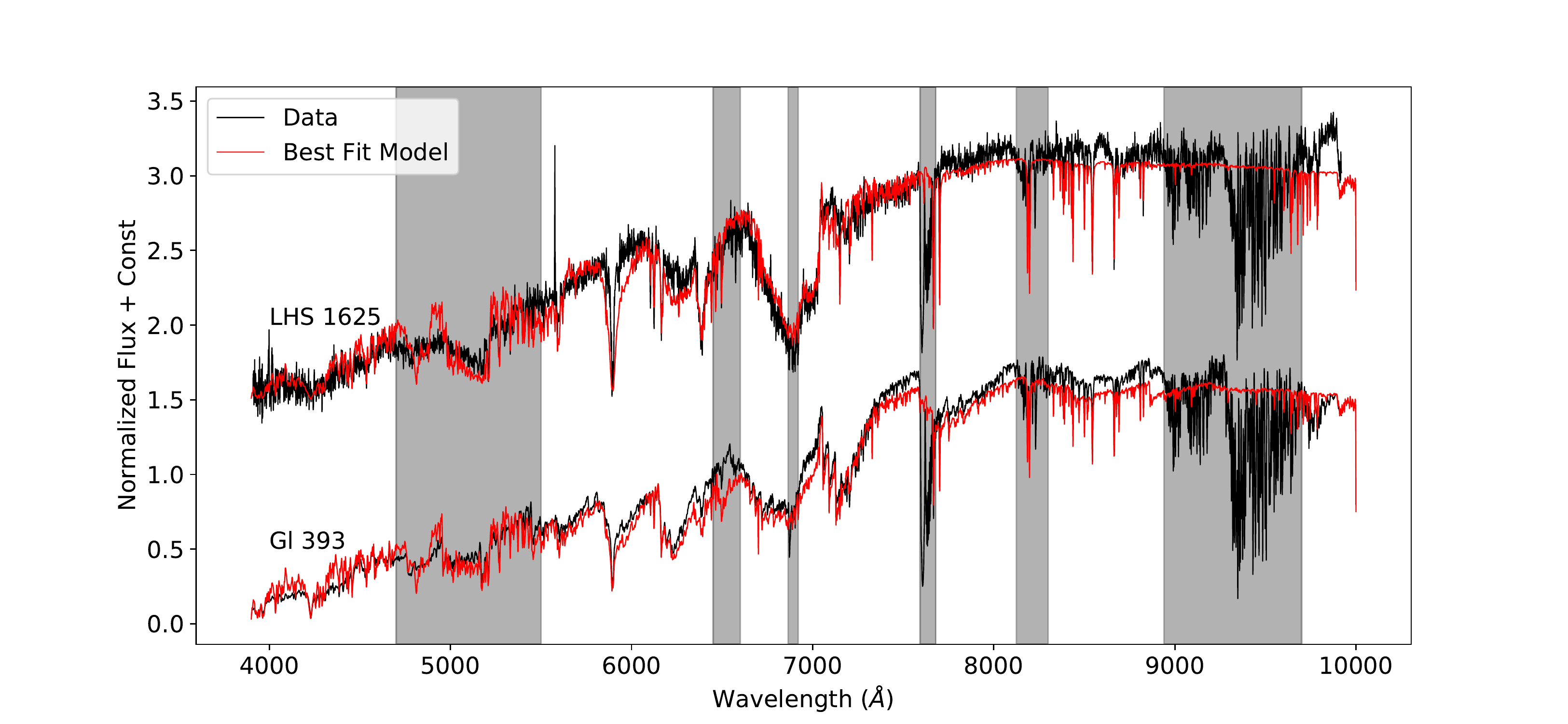}
\caption{\small
Example of two of our spectra (black) and their respective best fit model spectra (red). The gray regions are the regions that have their weights set to 0. The four regions red-ward of 6800 \AA\ are excluded due to telluric features. The region between $\sim$6400$-$6600\AA\ is a region where there is a known issue with a poorly modeled TiO absorption band \citep{reyle11}. The region around 5000\AA\ does not match the majority of our spectra (regardless of effective temperature or metallicity), and the scaling of the MgH band seems to be particularly problematic. LHS 1625 has a spectral type of usdM6, a best-fit effective temperature of 3400 K, a best-fit log($g$) of 5.5, and a [Fe/H] of $-$1.5. Gl 393 has a spectral type of dM2, a best-fit effective temperature of 3500 K, a best-fit log($g$) of 5.0, and a [Fe/H] of 0.0. }
\label{f:tempSpec}
\end{center}
\end{figure*}

To calculate the effective temperature we fit each spectrum to the BT-SETTL model grid using a method similar to that of \citet{mann13, mann15}. The BT-SETTL grid was created using the PHOENIX stellar atmosphere code \citep{allard12}. We chose to use the BT-SETTL grid that utilized the \citet{caffau11} solar abundances (CIFIST grid\footnote{\url{https://phoenix.ens-lyon.fr/Grids/BT-Settl/CIFIST2011/}}) since \citet{mann13} found that this grid of abundances gave the smallest errors in effective temperature when comparing model-fit effective temperature values to precisely known effective temperatures determined through long baseline optical interferometry.

The model grid we used was comprised of effective temperatures ranging from $2600-4500K$ in $100K$ bins, metallicities ranging from $-2.5$ to $+0.5$ dex in $0.5$ dex bins, and surface gravities (log $g$) of 4.5, 5.0, or 5.5 dex [$cm s^{-2}$]. This was the smallest-resolution grid publicly available for the CIFIST models.

To compare the models to an observed spectrum we convolved the models with a Gaussian kernel. We used the full width at half maximum (FWHM) of the spectrum and converted to the standard deviation ($\sigma \simeq \mathrm{FWHM} / 2.355$), which was then used as the standard deviations of the Gaussian kernel. We then determined a goodness-of-fit statistic ($G$) for each model $k$, given by the following equation from \citet{cushing08}:

\begin{equation}
    G_{k} = \sum_{i=1}^{n} \left( \frac{w_{i}(F_{i} - C_{k}\mathcal{F}_{i,k})}{\sigma_{i}}\right)^2
\end{equation}

where $n$ is the total number of data pixels, $w_{i}$ is a weight assigned to each data pixel, $F_{i}$ is the flux density of each data pixel, $\mathcal{F}_{i,k}$ is the flux density of each model $k$ pixel, $\sigma_{i}$ is the uncertainty in each data pixel, and $C_{k}$ is a normalization constant. For absolute flux calibrated stars, $C_{k}$ is equal to $R^2/D^2$; however, since $R$ is unknown, we followed \citet{mann13} and set this constant so that the mean of $F$ and $F_{k}$ were the same. The model spectrum chosen as the best fit (and therefore our effective temperature estimate) was the one which minimized the goodness-of-fit statistic ($G$).

The weights $w_{i}$ were set to either 0 or 1 so as to exclude regions in our spectra that were contaminated by telluric features, or regions where models did not accurately fit observed spectra of low-mass stars. These regions are shown with gray boxes in Figure \ref{f:tempSpec}. More details on which regions were excluded and why are given in the caption for Figure \ref{f:tempSpec}.

\begin{figure}[htp]
\begin{center}
\includegraphics[width=\linewidth]{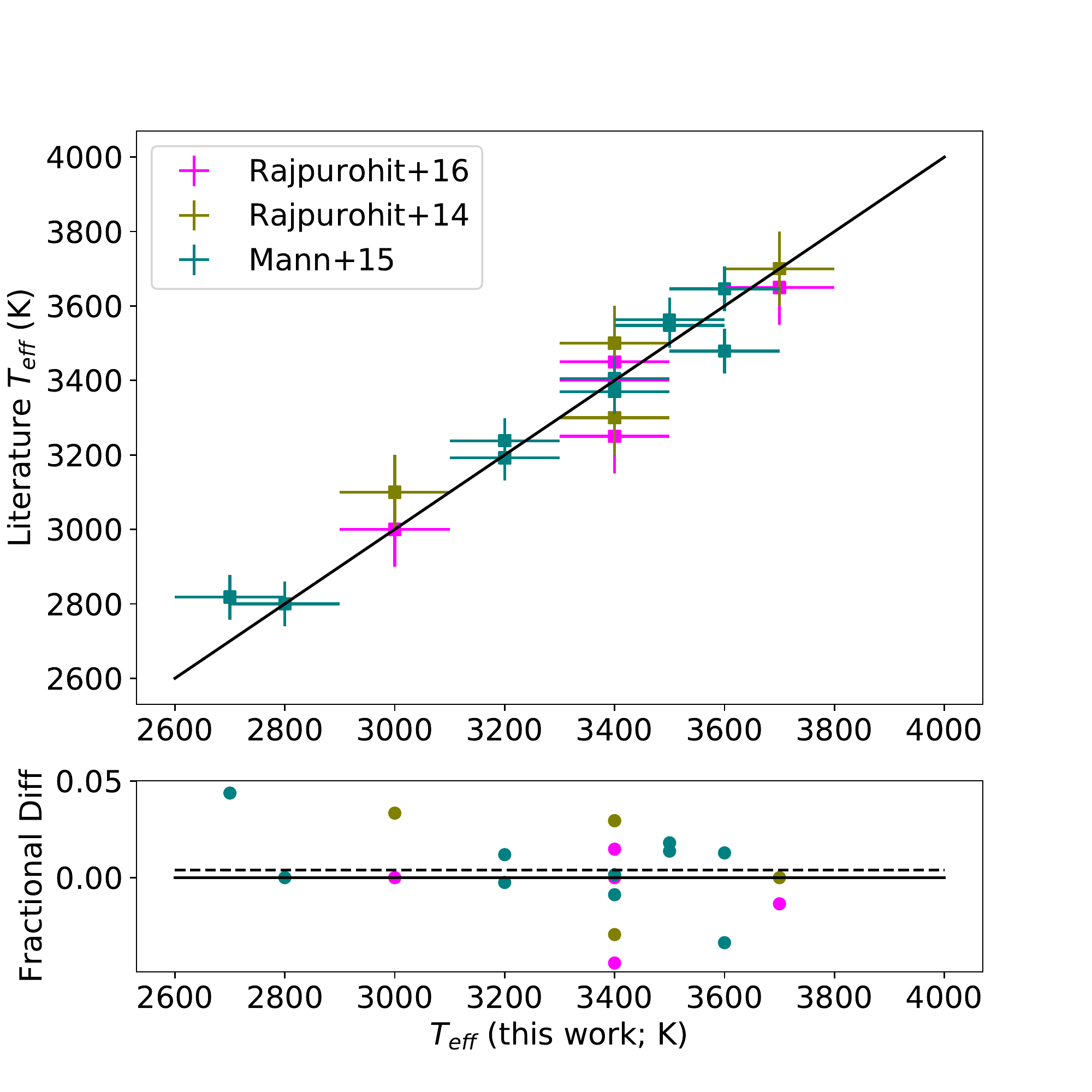}
\caption{\small
Comparison between our temperatures and those measured by previous studies. If our values and the literature values were exactly the same the fractional difference on the bottom plot would be exactly 0.0 (black solid line). The fractional difference is defined as the literature effective temperature minus our effective temperature divided by our effective temperature. We find a mean fractional difference of 0.3\% (dotted line). All of our effective temperatures deviate from previous literature values by 100 K or less except for one which deviates by 150 K. The 100 K mismatches seen between our values and those of \citet{rajpurohit14} are probably due to the coarse grid size (100 K) of both studies. }
\label{f:litTemp}
\end{center}
\end{figure}

To test the accuracy of our effective temperature measurements we compared them to the effective temperatures of stars in our sample that have previous literature values (Figure \ref{f:litTemp}). The technique in \citet{mann15} has been calibrated against effective temperatures derived using long baseline optical interferometry and shows typical uncertainties of 60 K, but does not contain subdwarf stars. Effective temperature estimates from \citet{rajpurohit14, rajpurohit16} measure the effective temperatures by fitting mid-to-high resolution optical and near-IR spectra to the same BT-Settl model grid as used here, but only measure effective temperatures for a small subset of M subdwarf stars. Our effective temperature estimates are consistent with all three previous literature effective temperature methods and show a mean fractional deviation of less than 1\%. We find that 83\% of our measurements fall within $1\sigma$ of the literature values and all of our measurements fall within $2\sigma$ of the literature measurements, leading us to conclude that our estimates are valid and accurate.

\begin{figure}[htp]
\begin{center}
\includegraphics[width=\linewidth]{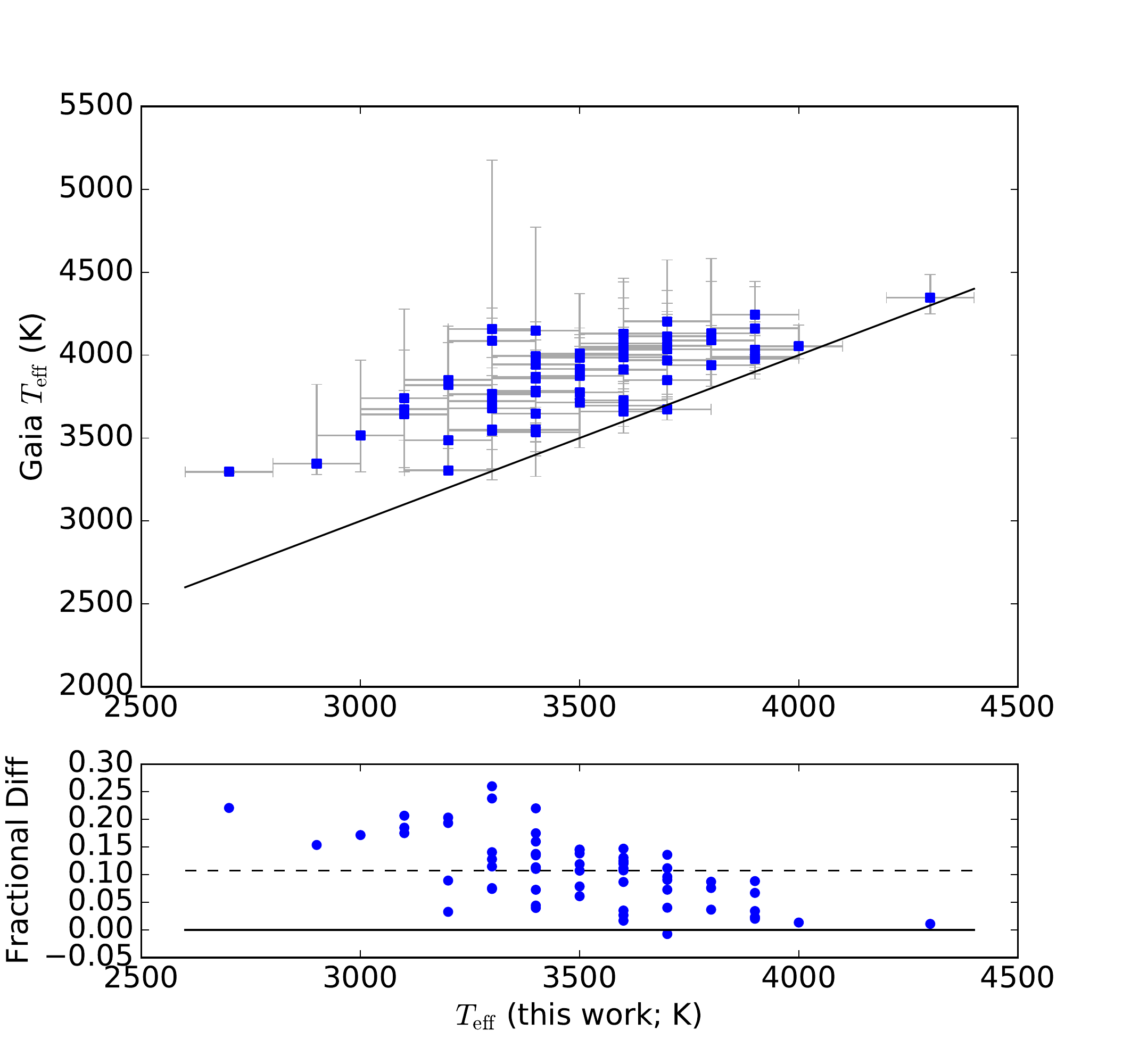}
\caption{\small
Comparison between our temperatures and those reported in Gaia DR2 \citep{andrae18}. Gaia overestimates the temperatures by a mean value of 10\%, however the temperatures below $\sim3200$ are overestimated by an even greater degree (almost 20\%). }
\label{f:gaiaTemp}
\end{center}
\end{figure}

We also compared our effective temperatures to those reported by Gaia DR 2 \citep{andrae18}. \citet{andrae18} use an empirically trained machine learning algorithm to determine a relation between Gaia $G$-, $R$-, and $B$-band photometry and previously determined $T_\mathrm{eff}$ measurements in the literature. We find that the effective temperatures listed in Gaia DR2 are higher than our effective temperatures by 10\% on average, and that the discrepancy is larger for cooler stars (see Figure \ref{f:gaiaTemp}). This discrepancy is not very surprising because the stars in our sample are at the edge of parameter space included in the machine learning training; the vast majority of the stars had near solar metallicities (95\% had [Fe/H] $> -$0.82 dex) and $T_\mathrm{eff}$ above 4000 K. Because of this, we do not use Gaia DR2 temperatures for any of our remaining analysis.

\subsection{Bolometric Luminosity}

\subsubsection{Compiling Photometry}
\label{s:phot}

We collected broad-band photometry for all of our sources, spanning the blue end of the optical region to mid-IR wavelengths. 
Optical photometry was collected from the Sloan Digital Sky Survey's $12^{th}$ data release \citep[SDSS; ][]{alam15}, the Pan-STARRS1 survey \citep{chambers16}, and from Gaia's Red and Blue Photometers \citep{gaia16}. All of the near-infrared (NIR) photometry was from the 2MASS All-Sky Point Source Catalog \citep[2MASS; ][]{skrutskie06}, with one source supplemented from the corresponding Reject Table (this source is noted in Table \ref{t:phot}). The \textit{Wide-Field Infrared Survey Explorer} \citep[WISE; ][]{wright10} AllWISE Point Source Catalog served as our source of mid-IR photometry. Both $WISE$ and 2MASS photometry were downloaded from IRSA\footnote{\url{http://irsa.ipac.caltech.edu/frontpage/}}. 

We imposed quality cuts to ensure that all the photometry was accurate, and examined each source by eye to ensure that there was no major background contamination. We only used SDSS photometry that had been flagged as ``clean", which selects the primary photometry for each source and rejects sources with any deblending problems, interpolation issues or saturation. The main issue with much of the Pan-STARRS photometry is the relatively high saturation limit, which is conservatively estimated to be 14.5, 15, 15, 14, and 13 for the $g$, $r$. $i$, $z$, and $y$ filters, respectively. Many fields are quoted to have reliable photometry up to a magnitude brighter than this, but to be conservative we chose to include only photometry brighter than these limits by at most a half magnitude, and only when there was no other indication of poor photometry (e.g., bad quality flags, or PSF did not include the entire source). For both $WISE$ and 2MASS data we did not include any photometry that was flagged as contaminated, saturated, or had a quality flag indicating that the photometry had a signal-to-noise ratio (SNR) less than five. We also visually inspected the $WISE$ W3 and W4 bands and did not include any photometry from these bands when the source was not visually discernible from the background. Since there are no quality flags for the Gaia DR2 data, we followed guidelines from \citet{evans18} and cut any sources with a color excess that exceeds $1.3 + 0.06(G_\mathrm{BP}$ - $G_\mathrm{RP})^2$, where $G_\mathrm{BP}$ is the Gaia blue-band magnitude and $G_\mathrm{RP}$ is the Gaia red-band magnitude. This relation removes any sources that have been affected by severe crowding, or calibration and processing issues. All of the final compiled photometry for each target is listed in Table \ref{t:phot}.  

\begin{center}
\begin{deluxetable*}{cccccccccccc}
\tabletypesize{\small}
\tablecaption{Photometry\label{t:phot}}
\tablehead{
\colhead{Star} &
\colhead{SDSS $u$} & 
\colhead{$\sigma_u$ } & 
\colhead{SDSS $g$} &
\colhead{$\sigma_g$} &  
\colhead{Pan-STARRS $g$} &  
\colhead{$\sigma_{PS1\ g}$ } &  
\colhead{Gaia $G_\mathrm{BP}$}  &
\colhead{$\sigma_{G_\mathrm{BP}}$ } &
\colhead{Pan-STARRS $r$} &
\colhead{$\sigma_{PS1\ r}$ } &
\colhead{...*}\\
\colhead{} &
\colhead{0.35 $\mu$m} &  
\colhead{} & 
\colhead{0.48 $\mu$m} &
\colhead{} & 
\colhead{0.481 $\mu$m} & 
\colhead{} & 
\colhead{0.5044 $\mu$m}  &
\colhead{} & 
\colhead{0.617 $\mu$m}  &
\colhead{} &
\colhead{}
}
\startdata
LHS1032&22.5&0.3&19.03&0.03&18.71&0.02&18.15&0.018&17.211&0.003\\ 
LHS104&17.06&0.01&14.48&0.02&--&--&13.969&0.001&--&--\\ 
LHS1074&24.1&1.1&20.18&0.02&19.84&0.02&19.25&0.06&18.374&0.004\\ 
LHS1166&22.4&0.3&19.99&0.02&19.64&0.01&19.22&0.06&18.247&0.003\\ 
LHS1174&21.1&0.1&18.03&0.02&17.81&0.006&17.28&0.01&16.378&0.004\\ 
LHS12&15.75&0.01&13.1950&0.0005&--&--&12.492&0.002&--&--\\ 
LHS1454&--&--&--&--&17.17&0.005&16.788&0.007&15.931&0.002\\ 
LHS156&--&--&--&--&15.651&0.001&15.205&0.003&--&--\\ 
LHS161&18.39&0.02&15.55&0.02&15.368&0.001&14.926&0.004&--&--\\ 
LHS1625&--&--&--&--&20.13&0.03&19.48&0.02&18.52&0.01\\ 
LHS1691&--&--&--&--&18.352&0.003&17.803&0.009&16.874&0.004\\ 
LHS170&--&--&--&--&--&--&10.891&0.001&--&--\\ 
LHS1703&17.82&0.03&15.18&0.04&14.846&3.0E-04&14.496&0.0020&--&--\\ 
\enddata
\tablecomments{*See online version or email the authors for full table, which includes all 85 objects and all photometry}
\end{deluxetable*}
\end{center}

Magnitudes were then converted to flux densities using the equation

\begin{equation}
\label{fluxEqn}
    F_\nu = F_{\nu 0} \times 10^{-m/2.5}
\end{equation}

where $F_\nu$ is the flux density, $m$ is the magnitude, and $F_{\nu 0}$ is the zero magnitude flux density. Gaia, 2MASS and $WISE$ magnitudes are given in the Vega photometric system, and $F_{\nu 0}$ is a constant that gives the same response as Vega for a given frequency ($\nu$). The zero magnitudes for 2MASS and $WISE$ were given in the Explanatory Supplements\footnote{\url{https://www.ipac.caltech.edu/2mass/releases/allsky/doc/sec6_4a.html}}$^,$\footnote{\url{http://wise2.ipac.caltech.edu/docs/release/allsky/expsup/sec4_4h.html}}, and for Gaia calculated using the Gaia B- and R-band filters and a model of Vega by the SVO Filter Profile Services\footnote{\url{http://svo2.cab.inta-csic.es/svo/theory/fps/index.php?mode=browse}} \citep{SVO1}. 
For $WISE$, we used the zero magnitudes derived using a constant power-law spectrum as recommended in the documentation since our sources were not steeply rising in the mid-IR. Pan-STARRS photometry is given in the AB magnitude system \citep{oke83} and thus has a constant zero magnitude flux for all bands. The SDSS magnitude system was intended to be an AB system, but is known to require slight adjustments \citep{fukugita96}, which are given in \citet{holberg06}. To then convert the SDSS magnitudes to fluxes we used the equations in the SDSS documentation\footnote{\url{http://www.sdss.org/dr12/algorithms/magnitudes/}} since the magnitudes are asinh magnitudes and not pogson magnitudes and Equation \ref{fluxEqn} cannot therefore be used. 

\begin{figure*}[ht]
\begin{center}
\includegraphics[scale=0.6]{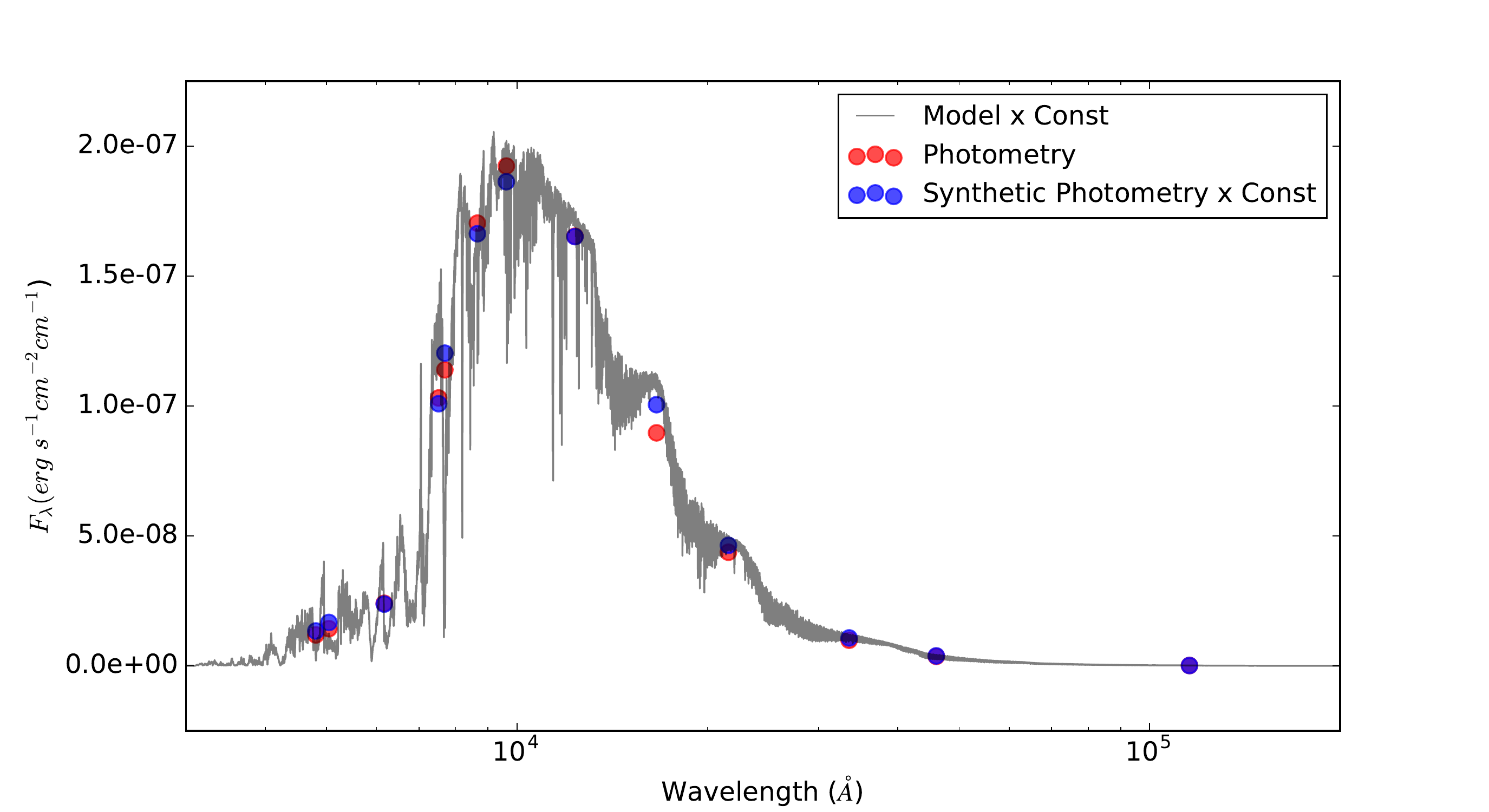}
\caption{\small
Spectral energy distribution (SED) of LHS 377. The red points show all the available photometry for the source, converted into $F_\lambda$  as described in Section \ref{s:phot}. The errorbars from the photometry are plotted but are similar in size to the points. The blue points are the synthetic photometry created using the filter bandpasses and gray model spectrum as described in Section \ref{s:Lbol}. The synthetic photometry and the model are both multiplied by the normalization constant $C$. The gray model multiplied by $C$ is what we integrate under to determine the bolometric flux and in turn the bolometric luminosity. }
\label{f:SED}
\end{center}
\end{figure*}

We converted $F_\nu$ to $F_\lambda$ using $F_\lambda = F_\nu (c / \lambda_{c}^2)$, where $c$ is the speed of light and $\lambda_c$ if the center of each filter bandpass, and given in Table \ref{t:phot}. These final values of $F_\lambda$ are the photometry values shown in Figure \ref{f:SED} and what we used for the remainder of the calculations involving photometry.

\subsubsection{The Bolometric Luminosity}
\label{s:Lbol}

Once the photometry was converted to physical flux densities, we used these points to anchor a spectrum. We chose to use the BT-SETTL model spectra throughout, since the flux calibration of the blue end of our spectra has known issues and there are large telluric absorption features contaminating the red side of our spectra. The best-fit BT-SETTL model from our effective temperature estimates (see Section \ref{s:Teff}) was normalized to fit the photometry. The normalization constant was determined by generating synthetic photometry from the model spectrum in a method similar to that of \citet{filippazzo15}. The synthetic photometry was generated from the best-fit model spectrum using filter transmission curves from the SVO Filter Profile Services and the following equation 

\begin{equation}
    F_{\lambda, synth} = \frac{\int_{}^{}T(\lambda)F_{\lambda, model}(\lambda)d\lambda}{\int_{}^{}T(\lambda)d\lambda}
\end{equation}

where $T(\lambda)$ is the transmission curve from SVO, interpolated onto the same wavelength grid as the model spectrum ($F_{\lambda, model}$). The normalization constant was found by then minimizing the squared difference between the synthetic and catalog photometry. The optimal minimization (and hence value of the normalization constant) was determined using the scipy routine \texttt{scipy.optimize.minimize$\_$scalar}. 

The bolometric luminosity was then determined by the following integral 

\begin{equation}
    L_\mathrm{bol} = 4 \pi D^2 \int^{500\ \mu m}_{0.1\ \mu m} C \times F_\lambda d\lambda
\end{equation}

where $C$ is the above determined normalization constant, $F_\lambda$ is the model flux, and $D$ is the distance determined from Gaia DR2 parallaxes \citep{gaia18}. Instead of using the inverted parallax to get $D$, we used the distances reported by \citet{bailer18} for Gaia DR2, which are publicly available within the Gaia archive external catalog \texttt{external.gaiadr2$\_$geometric$\_$distance}. The \citet{bailer18} distances are more reliable because they account for the nonlinearity of the transformation from parallax to distance. This nonlinearity is corrected using a Bayesian distance prior that varies as a function of galactic longitude and latitude. Finally, we used a simple trapezoidal integration (\texttt{numpy.trapz}) to numerically integrate $F_\lambda$ over the stated wavelength range. 

To determine how the model parameters ($T_\mathrm{eff}$, [Fe/H], and log $g$) influenced our bolometric luminosity calculation, we tested how varying these parameters changed our estimate of $L_\mathrm{bol}$. We found that by changing the model by one grid point, $\log_{10}(L_\mathrm{bol}/L_\mathrm{Sun})$ changed by an average of $0.008\pm0.005$, $0.007\pm0.005$, and $0.002\pm0.002$ for a change in $T_\mathrm{eff}$ of 100K, and [Fe/H] and log $g$ of 0.5 dex, respectively. If all three are changed in conjunction, the change in $\log_{10}(L_\mathrm{bol}/L_\mathrm{Sun})$ was on average $0.015\pm0.008$, however we do not expect our estimates to deviate this substantially in all three parameters. These errors are larger than the propagated uncertainties, and so we adopt the change if all three parameters are changed in conjunction as a conservative estimate of the uncertainty in the bolometric flux (the uncertainty in the parallax is then incorporated to determine the total uncertainty in $L_\mathrm{bol}$).

We also compared how using real spectra versus models changed our values of $L_\mathrm{bol}$. Three of our targets had previously published spectra that spanned the near- and mid-IR (LHS 1174, LHS 377, LSR J2122+36, all from the SpeX Prism Spectral Libraries \citep{burgasser14}\footnote{\url{http://pono.ucsd.edu/~adam/browndwarfs/spexprism/}}). In combination with our optical spectra, a majority of the flux-contributing region of the SED was covered by real spectra. We found that by using the real spectra instead of the best-fit model, $log_{10}(L_{bol}/L_{Sun})$ changed by 0.01. This value is well within our new adopted uncertainties from changing the model, so we conclude that using a model instead of a real spectrum is indeed valid (as long as the uncertainties mentioned above are included).

\section{Results}
\label{s:Results}

\begin{figure*}[ht]
\begin{center}
\includegraphics[width=\linewidth]{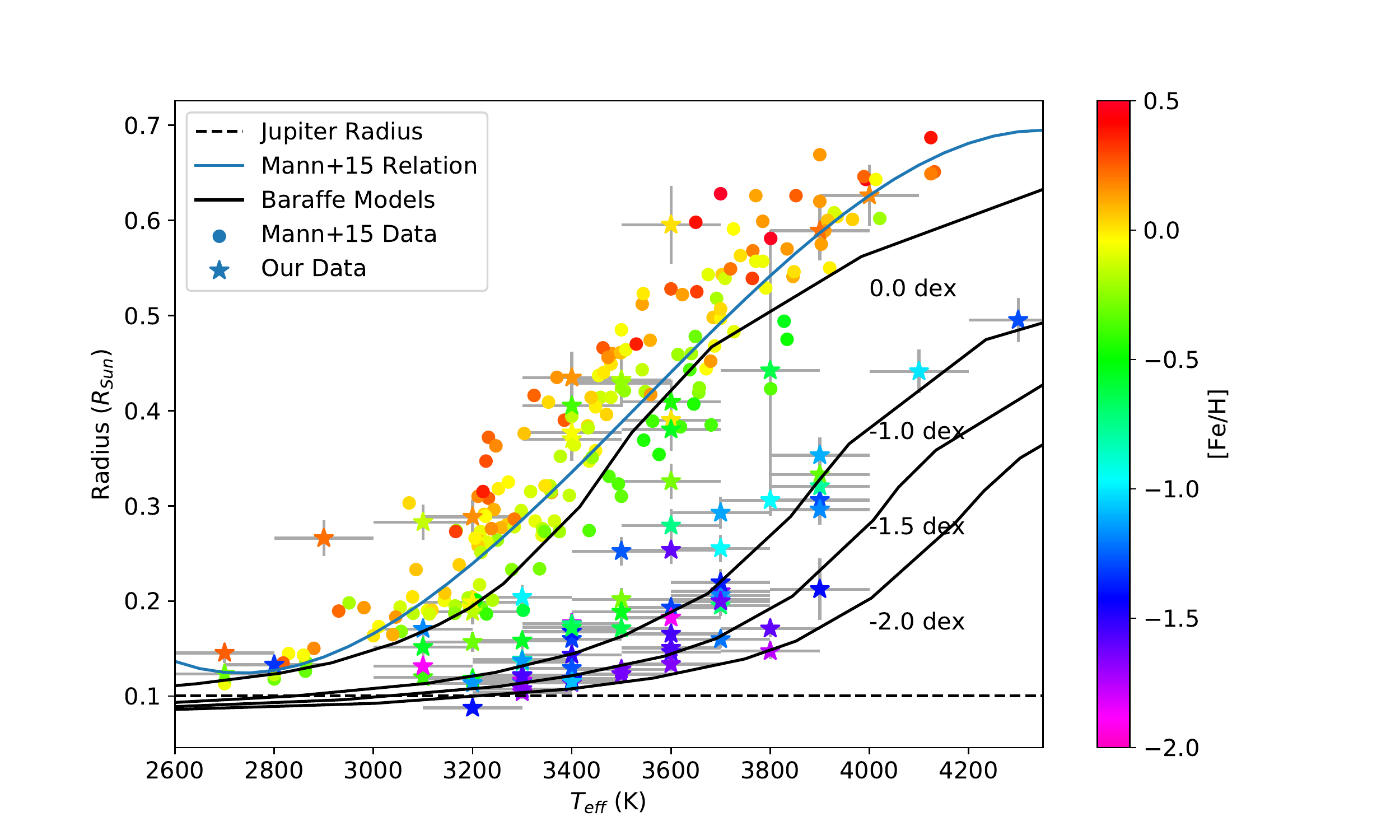}
\caption{\small
Results of our effective temperature and radius determinations of all the stars in our sample (star markers), as well as previously determined effective temperatures and radii from \citet{mann15} (circle markers). The points are colored by their metallicity ([Fe/H]). The empirically determined relation from \citet{mann15} for solar metallicity stars is shown as a blue line, while the relations from the Baraffe stellar evolutionary models \citep{baraffe97} are shown in black. Our points fall along the stellar evolution curves and thus validate the predicted factor of two to three change in radius for extreme and ultra subdwarfs for a given effective temperature.}
\label{f:TeffVsRad}
\end{center}
\end{figure*}

The effective temperatures (calculated in Section \ref{s:Teff}) and bolometric luminosities (calculated in Section \ref{s:Lbol}) were combined to determine a radius using the Stefan-Boltzmann Law: $R = \sqrt{L_\mathrm{bol}/ (4\pi \sigma T_\mathrm{eff}^4)}$. The derived parameters (including $T_\mathrm{eff}$, $L_\mathrm{bol}$, and $R$) for all of our sources are given in Table \ref{t:params}. Figure \ref{f:TeffVsRad} shows how the radii change for a given effective temperature with decreasing metallicity. We find that stellar evolutionary models from \citet{baraffe97} accurately predict the radii of low-metallicity subdwarfs. For a given effective temperature the radius can deviate by a factor of almost five for a change in metallicity of 2.5 dex.

\begin{center}
\begin{longtable*}{lllllllllll}
\caption{Derived Parameters \label{t:params}} \\
\toprule
Star & Spectral & $T_\mathrm{eff}$ & $\sigma_{T}$ & $log(L_\mathrm{bol}/L_\mathrm{Sun})$ & $\sigma_{\log(L_\mathrm{bol}/L_\mathrm{Sun})}$ & Radius & $\sigma_R$ & [Fe/H] & $\sigma_{[Fe/H]}$ & [Fe/H] \\
Name & Class & (K) & (K) & & & ($R_\mathrm{Sun}$) & ($R_\mathrm{Sun}$) & & & method \\ \midrule
\endfirsthead
\toprule
Star & Spectral & $T_\mathrm{eff}$ & $\sigma_{T}$ & $\log(L_\mathrm{bol}/L_\mathrm{Sun})$ & $\sigma_{\log(L_\mathrm{bol}/L_\mathrm{Sun})}$ & Radius & $\sigma_R$ & [Fe/H] & $\sigma_{[Fe/H]}$ & [Fe/H] \\
 Name & Class & (K) & (K) & & & ($R_\mathrm{Sun}$) & ($R_\mathrm{Sun}$) & & & method\\ \midrule
\endhead
\hline
\multicolumn{11}{c}{Continued}\\   \bottomrule
\endfoot
\bottomrule
\endlastfoot
2MASSJ0822+1700 & usdM6 & 3200 & 100 & $-$3.139 & 0.031 & 0.088 & 0.006&$-$1.4& 0.3 & Spec\\ 
Gl109 & dM3 & 3400 & 100 & $-$1.783 & 0.058 & 0.37 & 0.033 &$-$0.1& 0.08 & Lit\footnote[1]{\citet{mann15} \label{mann}}\\ 
Gl143.1 & dK7 &4000 & 100 & $-$1.044 & 0.011 & 0.626 & 0.033&0.17& 0.15 & Phot \\ 
Gl229A & dM1 & 3600.0 & 100 & $-$1.271 & 0.035 & 0.595 & 0.041 & 0.02 & 0.08 & Lit$^{\ref{mann}}$ \\
Gl270 & dM0 &3900& 100 & $-$1.141 & 0.011 & 0.589 & 0.03&0.23& 0.15 & Phot \\ 
Gl388 & dM3 &3400 & 100 & $-$1.643 & 0.02 & 0.435 & 0.027 &0.15& 0.08 & Lit$^{\ref{mann}}$ \\ 
Gl393 & dM2 &3500 & 100 & $-$1.597 & 0.01 & 0.432 & 0.025&$-$0.18& 0.08 & Lit$^{\ref{mann}}$ \\ 
Gl402 & dM4 &3200 & 100 & $-$2.105 & 0.013 & 0.288 & 0.019 &0.16& 0.08 & Lit$^{\ref{mann}}$ \\ 
Gl406 & dM6 &2700 & 100 & $-$2.995 & 0.007 & 0.145 & 0.011&0.25& 0.08 & Lit$^{\ref{mann}}$ \\ 
Gl411 & dM2 & 3400 & 100 & $-$1.704 & 0.037 & 0.405 & 0.029&$-$0.38& 0.08 & Lit$^{\ref{mann}}$\\
Gl436 & dM3 &3600 & 100 & $-$1.638 & 0.015 & 0.39 & 0.023&0.01& 0.08 & Lit$^{\ref{mann}}$ \\ 
Gl447 & dM4 &3200 & 100 & $-$2.43 & 0.014 & 0.198 & 0.013&$-$0.02& 0.08 & Lit$^{\ref{mann}}$\\ 
Gl51 & dM5 &2900 & 100 & $-$2.346 & 0.013 & 0.266 & 0.019&0.22& 0.08 & Lit\footnote[2]{\citet{gaidos14}} \\ 
Gl908 & dM1 &3600 & 100 & $-$1.596 & 0.011 & 0.409 & 0.023&$-$0.45& 0.08 & Lit$^{\ref{mann}}$ \\ 
LHS1032 & usdM4 &3400 & 100 & $-$2.775 & 0.02 & 0.118 & 0.007&$-$1.4& 0.3 & Spec \\ 
LHS104 & esdK7 &3900 & 100 & $-$1.711 & 0.006 & 0.306 & 0.016& $-$1.29 & 0.3& Spec\\ 
LHS1074 & sdM6 &3200 & 100 & $-$2.88 & 0.028 & 0.118 & 0.008& $-$0.52& 0.3 & Spec\\ 
LHS1166 & sdM6 &3100 & 100 & $-$2.924 & 0.024 & 0.12 & 0.008&$-$0.39& 0.3 & Spec \\ 
LHS1174 & esdM3 &3400 & 100 & $-$2.513 & 0.013 & 0.16 & 0.01&$-$1.31& 0.3 & Spec \\ 
LHS12 & d/sdM0 &3900 & 100 & $-$1.642 & 0.019 & 0.331 & 0.018&$-$0.33& 0.15 & Phot \\ 
LHS1454 & usdK7 &3800 & 100 & $-$2.262 & 0.012 & 0.171 & 0.009&$-$1.59& 0.3 & Spec \\ 
LHS156 & sdM3 &3500 & 100 & $-$2.403 & 0.009 & 0.171 & 0.01&$-$0.64& 0.3 & Spec\\ 
LHS161 & esdM2 &3700 & 100 & $-$2.166 & 0.006 & 0.201 & 0.011&$-$1.1& 0.3 & Spec \\ 
LHS1625 & usdM6 &3400 & 100 & $-$2.809 & 0.041 & 0.114 & 0.009&$-$1.64& 0.3 & Spec \\ 
LHS1691 & usdM2 &3400 & 100 & $-$2.429 & 0.014 & 0.176 & 0.011&$-$1.8& 0.3 & Spec \\ 
LHS170 & esdK7 &4300 & 100 & $-$1.123 & 0.008 & 0.495 & 0.023&$-$1.28& 0.3 & Spec \\ 
LHS1703 & esdK7 & 3900 & 100 & $-$1.587 & 0.012 & 0.352 & 0.019 & $-$1.1 & 0.3 & Spec \\
LHS173 & esdK7 &4100 & 100 & $-$1.305 & 0.016 & 0.441 & 0.023&$-$0.94& 0.18 & Lit\footnote[2]{\citet{schmidt16}} \\ 
LHS174 & sdM0 &3800 & 100 & $-$1.434 & 0.32 & 0.442 & 0.165&$-$0.63& 0.3 & Spec \\ 
LHS1742a & esdM6 &3300 & 100 & $-$2.912 & 0.333 & 0.107 & 0.042 &$-$0.97& 0.3 & Spec \\ 
LHS178 & d/sdM1 &3600 & 100 & $-$1.795 & 0.013 & 0.326 & 0.019&$-$0.29& 0.3 & Spec \\ 
LHS182 & usdM0 &3700 & 100 & $-$2.128 & 0.085 & 0.21 & 0.024&$-$1.66& 0.3 & Spec \\ 
LHS1826 & usdM6 &3300 & 100 & $-$2.94 & 0.019 & 0.104 & 0.007&$-$1.73& 0.3 & Spec \\ 
LHS1863 & usdM1 &3600 & 100 & $-$2.015 & 0.01 & 0.253 & 0.014&$-$1.59& 0.3 & Spec \\ 
LHS1994 & esdM1 &3700 & 100 & $-$1.844 & 0.017 & 0.291 & 0.017&$-$1.13& 0.3 & Spec \\ 
LHS20 & d/sdM2 &3500 & 100 & $-$2.26 & 0.011 & 0.202 & 0.012&$-$0.28& 0.15 & Spec \\ 
LHS2023 & esdM6 &3200 & 100 & $-$2.917 & 0.022 & 0.113 & 0.008&$-$1.15& 0.3 & Spec \\ 
LHS205a & usdM5 &3400 & 100 & $-$2.783 & 0.028 & 0.117 & 0.008&$-$1.43& 0.3 & Spec \\ 
LHS2061 & sdM5 &3300 & 100 & $-$2.691 & 0.019 & 0.138 & 0.009&$-$0.76& 0.3& Spec \\ 
LHS2096 & esdM5 &3300 & 100 & $-$2.852 & 0.018 & 0.115 & 0.007&$-$1.25& 0.3 & Spec \\ 
LHS2163 & sdM1 &3600 & 100 & $-$1.661 & 0.017 & 0.38 & 0.022&$-$0.56& 0.12 & iSHELL Spec \\ 
LHS228 & sdM2 &3500 & 100 & $-$2.32& 0.019 & 0.188 & 0.012&$-$0.55& 0.3 & Spec \\ 
LHS2326 & esdM2 &3300 & 100 & $-$2.353 & 0.009 & 0.204 & 0.013&$-$0.98& 0.3 & Spec \\ 
LHS2355 & usdM0 &3800 & 100 & $-$2.393 & 0.014 & 0.147 & 0.008&$-$1.76& 0.3 & Spec \\ 
LHS2405 & d/sdM4 &3500 & 100 & $-$1.604 & 0.011 & 0.429 & 0.025&$-$0.24& 0.15& Spec \\ 
LHS2500 & usdM5 &3100 & 100 & $-$2.845 & 0.039 & 0.131 & 0.01&$-$1.88& 0.3 & Spec \\ 
LHS2674 & sdM4 &3300 & 100 & $-$2.573 & 0.022 & 0.158 & 0.01&$-$0.57 & 0.3& Spec \\ 
LHS272 & sdM3 &3400 & 100 & $-$2.431 & 0.01 & 0.175 & 0.011&$-$0.72& 0.3& Spec\\
LHS2843 & esdM0 &3500 & 100 & $-$2.068 & 0.015 & 0.251 & 0.015&$-$1.26& 0.3 & Spec \\ 
LHS2852 & sdM2 &3400 & 100 & $-$1.767 & 0.01 & 0.377 & 0.023&$-$0.05& 0.12 & iSHELL Spec \\ 
LHS3090 & usdM4 &3400 & 100 & $-$2.609 & 0.015 & 0.143 & 0.009&$-$1.5& 0.3 & Spec \\ 
LHS318 & esdM1 &3600 & 100 & $-$2.25 & 0.01 & 0.193 & 0.011&$-$1.3& 0.3 & Spec \\ 
LHS3189 &d/sdM1 &3100 & 100 & $-$2.72 & 0.022 & 0.151 & 0.01&$-$0.57& 0.15 & Phot\\ 
LHS3255 & dM4 &3100 & 100 & $-$2.177 & 0.009 & 0.283 & 0.018&$-$0.15& 0.15 & Phot\\ 
LHS326 & esdM3 &3700 & 100 & $-$2.147 & 0.007 & 0.206 & 0.011&$-$1.18& 0.3 & Spec \\ 
LHS3263 & esdM3 &3700 & 100 & $-$2.369 & 0.019 & 0.159 & 0.009&$-$1.22& 0.3 & Spec \\ 
LHS3276 & esdK7 &3900 & 100 & $-$1.741 & 0.014 & 0.295 & 0.016&$-$1.18& 0.3& Spec \\ 
LHS3382 & usdM3 &3400 & 100 & $-$2.472 & 0.013 & 0.167 & 0.01&$-$1.38& 0.3 & Spec \\ 
LHS3390 & sdM5 &3300 & 100 & $-$2.708 & 0.014 & 0.135 & 0.008&$-$0.83& 0.3 & Spec \\ 
LHS3409 & d/sdM4 &3200 & 100 & $-$2.635 & 0.019 & 0.157 & 0.01&$-$0.31& 0.12 & iSHELL Spec \\ 
LHS3555 & usdM2 &3300 & 100 & $-$2.842 & 0.022 & 0.116 & 0.008&$-$1.78& 0.3 & Spec \\ 
LHS360 & esdM0 &3700 & 100 & $-$1.96 & 0.013 & 0.255 & 0.014&$-$0.96& 0.3 & Spec \\ 
LHS364 & usdM1 &3600 & 100 & $-$2.491 & 0.014 & 0.146 & 0.008&$-$1.54& 0.3 & Spec \\ 
LHS375 & esdM4 &3400 & 100 & $-$2.697 & 0.01 & 0.129 & 0.008&$-$1.27& 0.3 & Spec \\ 
LHS377 & sdM7 &3000 & 100 & $-$2.993 & 0.019 & 0.118 & 0.008&$-$0.41 & 0.3 & Spec\\ 
LHS4028 & usdM4 & 3500 & 100 & $-$2.692 & 0.018 & 0.123 & 0.007&$-$1.64& 0.3 & Spec\\ 
LHS42 & esdM0 &3800 & 100 & $-$1.756 & 0.008 & 0.306 & 0.016&$-$0.96& 0.12 & iSHELL Spec \\ 
LHS453 & usdM4 &3300 & 100 & $-$2.799 & 0.026 & 0.122 & 0.008&$-$1.77& 0.3 & Spec\\ 
LHS482 & sdM1 & 3600 & 100 & $-$1.929 & 0.026 & 0.279 & 0.018&$-$0.75& 0.12 & iSHELL Spec \\
LHS489 & usdM0 &3600 & 100 & $-$2.299 & 0.017 & 0.182 & 0.011&$-$1.88& 0.3 & Spec\\ 
LHS504 & d/sdM5 &3100 & 100 & $-$2.588 & 0.026 & 0.176 & 0.012&$-$0.18& 0.3 & Spec \\ 
LHS515 & esdM5 &3400 & 100 & $-$2.8 & 0.014 & 0.115 & 0.007&$-$1.08& 0.3 & Spec \\ 
LHS518 & sdK7 &3900 & 100 & $-$1.671 & 0.018 & 0.32 & 0.018&$-$0.79& 0.3 & Spec\\ 
LHS522 & usdK7 &3900 & 100 & $-$2.027 & 0.127 & 0.212 & 0.033&$-$1.41& 0.3 & Spec\\ 
LSRJ0020+5526 & sdM2 &3700 & 100 & $-$2.194 & 0.015 & 0.195 & 0.011&$-$0.7& 0.3 & Spec \\ 
LSRJ0522+3814 & usdM3 &3500 & 100 & $-$2.655 & 0.01 & 0.128 & 0.007&$-$1.63& 0.3 & Spec \\ 
LSRJ0621+1219 & usdM6 &3300 & 100 & $-$2.912 & 0.014 & 0.107 & 0.007&$-$1.65& 0.3 & Spec\\ 
LSRJ0621+3652 & usdK7 &3700 & 100 & $-$2.091 & 0.008 & 0.219 & 0.012&$-$1.38& 0.3 & Spec \\ 
LSRJ0705+0506 & sdM4 &3400 & 100 & $-$2.451 & 0.013 & 0.171 & 0.01&$-$0.64& 0.15 & Phot \\ 
LSRJ1340+1902 & esdM4 &3300 & 100 & $-$2.698 & 0.016 & 0.137 & 0.009&$-$1.15& 0.3 & Spec \\ 
LSRJ1956+4428 & usdM0 &3600 & 100 & $-$2.465 & 0.008 & 0.15 & 0.008&$-$1.56& 0.3 & Spec \\ 
LSRJ2115+3804 & usdK7 &3700 & 100 & $-$2.174 & 0.007 & 0.199 & 0.011&$-$1.62& 0.3 & Spec \\ 
LSRJ2122+3656 & esdM5 &3300 & 100 & $-$2.802 & 0.011 & 0.122 & 0.008&$-$1.34& 0.3 & Spec\\ 
LSRJ2205+5353 & usdM1 &3600 & 100 & $-$2.384 & 0.009 & 0.165 & 0.009&$-$1.55& 0.3 & Spec \\ 
NLTT3247 & dM4 & 3200 &100 & $-$2.475 & 0.026 & 0.188 & 0.013&$-$0.09& 0.15 & Phot \\ 
Teegarden & dM6 & 2700 &100 & $-$3.137 & 0.001 & 0.123 & 0.009&$-$0.31& 0.08 & Lit$^{\ref{mann}}$ \\ 
WISE0238+3617 & usdM3 & 3300 &100 & $-$2.807 & 0.015 & 0.121 & 0.008&$-$1.56& 0.3 & Spec \\ 
WISE0707+1705 & usdM2 & 3600 &100 & $-$2.57 & 0.012 & 0.133 & 0.008&$-$1.65& 0.3 & Spec \\
\end{longtable*}
\end{center}

\subsection{Color Relations}

\begin{figure*}[ht]
\begin{center}
\includegraphics[width=\linewidth]{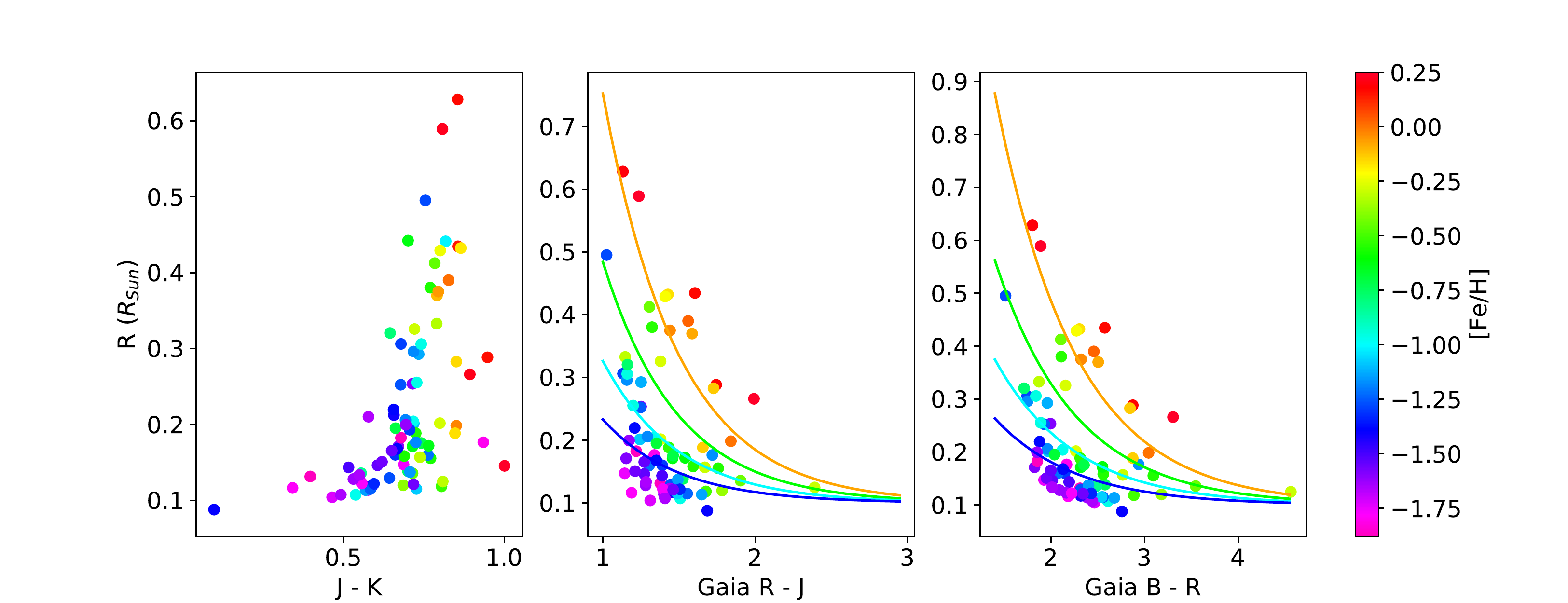}
\caption{\small
2MASS and Gaia broadband colors versus stellar radius. Stars with similar colors show large variations in radius for different metallicities. Overplotted on the two plots on the right are our color$-$Radius relations for metallicity values of 0.0 (orange), $-$0.5 (green), $-$1.0 (cyan), and $-$1.5 (blue). Even with these metallicity dependent relations we find a 1$-\sigma$ scatter of $\sim20\%$ in the radius. The fits are given in Equation \ref{e:colorR}. }
\label{f:ColorR}
\end{center}
\end{figure*}

Broadband colors are readily available for a massive number of sources thanks to surveys such as Gaia and 2MASS. We therefore present Gaia and 2MASS color to radius and absolute magnitude relations for our sources. Figure \ref{f:ColorR} shows different optical and IR color to radius relations. 
We find that $J-K$ is not well fit by a simple equation, but both Gaia $R - J$ and Gaia $R- B$ can be fit with equations relating these colors to the stellar radius. We chose a decreasing exponential equation to describe the data, which was physically motivated by the fact that the stellar radii cannot collapse to sizes smaller than $0.1 R_\mathrm{Sun}$ due to degeneracy pressure. We use the following exponential to describe the data: 

\begin{equation}
R = A\ e^{ - [b(color) + c[Fe/H]]}
\label{e:colorR}
\end{equation}

\noindent where the best fit constants for Gaia $R - J$ are 5.02, 2.04, and -1.06 and for Gaia $B - R$  are 4.0, 1.17, and $-$1.04 for $A$, $b$, and $c$, respectively. Even with a metallicity dependent relation we still find a scatter in the radius of $\sim20\%$. 

We also fit color to metallicity relations for our sample. Like previous studies \citep[e.g., ][]{mann13, newton14, mann15} we find that $J - K_{s}$ gives the best fit for a single color to [Fe/H] relation, and find the following best-fit equation: 

\begin{equation}
[Fe/H] = 4.22(J - K_{s}) - 3.86
\end{equation}

\noindent where the 1-$\sigma$ scatter is 0.37 dex. 

\subsection{Absolute Magnitude Relations}

\begin{figure}[ht]
\begin{center}
\includegraphics[width=\linewidth]{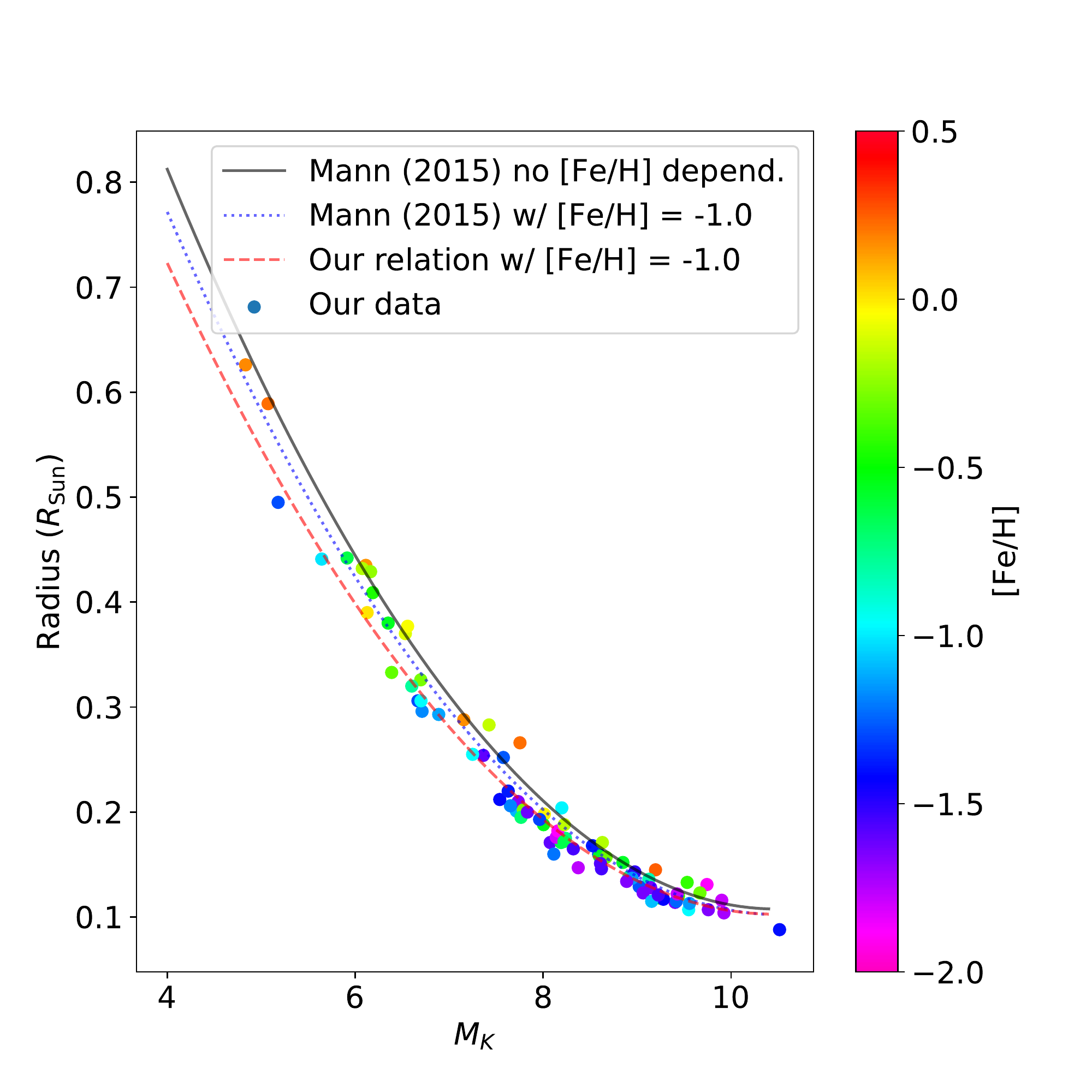}
\caption{\small
Absolute $K_s$-band versus radius relation for our entire sample of stars. In black is the best fit relation from \citet{mann15}, which is valid for stars with $[Fe/H] > -0.6$ and does not include metallicity as a parameter. In blue, we plot the metallicity dependent relation, which has the form of Equation \ref{e:MkRad}, extrapolated past its tested metallicity limit ($-$0.6 dex) at a value of $-$1.0 dex. We find that while this better fits our data, it still over-predicts the radii of the lowest metallicity stars in our sample. In red, we plot our new metallicity dependent relation at a value of $-$1.0 dex.}
\label{f:MkRad}
\end{center}
\end{figure}

Previous studies have found that the scatter in radius relations due to metallicity can be reduced (or even eliminated) by using absolute infrared photometry versus radius relations \citep[$M_{K_s} -$ Radius: e.g., ][]{boyajian12, mann15}. However, the spread in metallicity previously explored was only about 1.0 dex (from $+0.5$ to $-0.5$ dex). We calculate absolute K-band magnitudes for our whole sample and find that while there is significantly less scatter for radii determined using an $M_{K_s} -$ Radius relation, the relation is still metallicity dependent (see Figure \ref{f:MkRad}).  For our lowest metallicity stars ($[Fe/H] < -1.0$ dex), we measure radii that are on average 10\% smaller than the radii that would be determined using the $M_{K} -$ Radius relation that does not include metallicity as a parameter \citep[Equation 4: ][]{mann15}. Equation 5 of \citet{mann15} gives a relation that includes metallicity as a parameter:

\begin{equation} 
R = (a - b M_{K_s} + c M_{K_s}^2) \times (1+ f[Fe/H])
\label{e:MkRad}
\end{equation} 

where they find best fit values of 1.9305, -0.3466, 0.01647, and 0.04458 for $a,b,c$, and $f$, respectively. We find that this relation fits our data better, but still overestimates the radii of our sample by an average of 5\% for stars with metallicities below -0.5 dex. We use our data to determine new coefficients that are valid for [Fe/H] values down to -2.0 dex, and find values of $1.875\pm 0.05$, $-0.337\pm 0.01$, $0.0161 \pm 0.0009$, and $0.079 \pm 0.01$ for $a,b,c$, and $f$, respectively. The scatter in the residuals of our $M_{K_s} -$ Radius relation is 6\% and is valid for $M_{K_s}$ values of 4 to 11 and metallicities from +0.5 dex to $-$2.0 dex. 

The absolute $K_{s}$-band relation greatly reduces the uncertainty in the radius compared to the color$-$radius relation (Equation \ref{e:colorR}) and so we recommend using it to get the more accurate radii whenever possible.

\subsection{Color Relations Relevant For \textit{WFIRST} Microlensing}

\begin{figure*}[ht]
\begin{center}
\includegraphics[width=\linewidth]{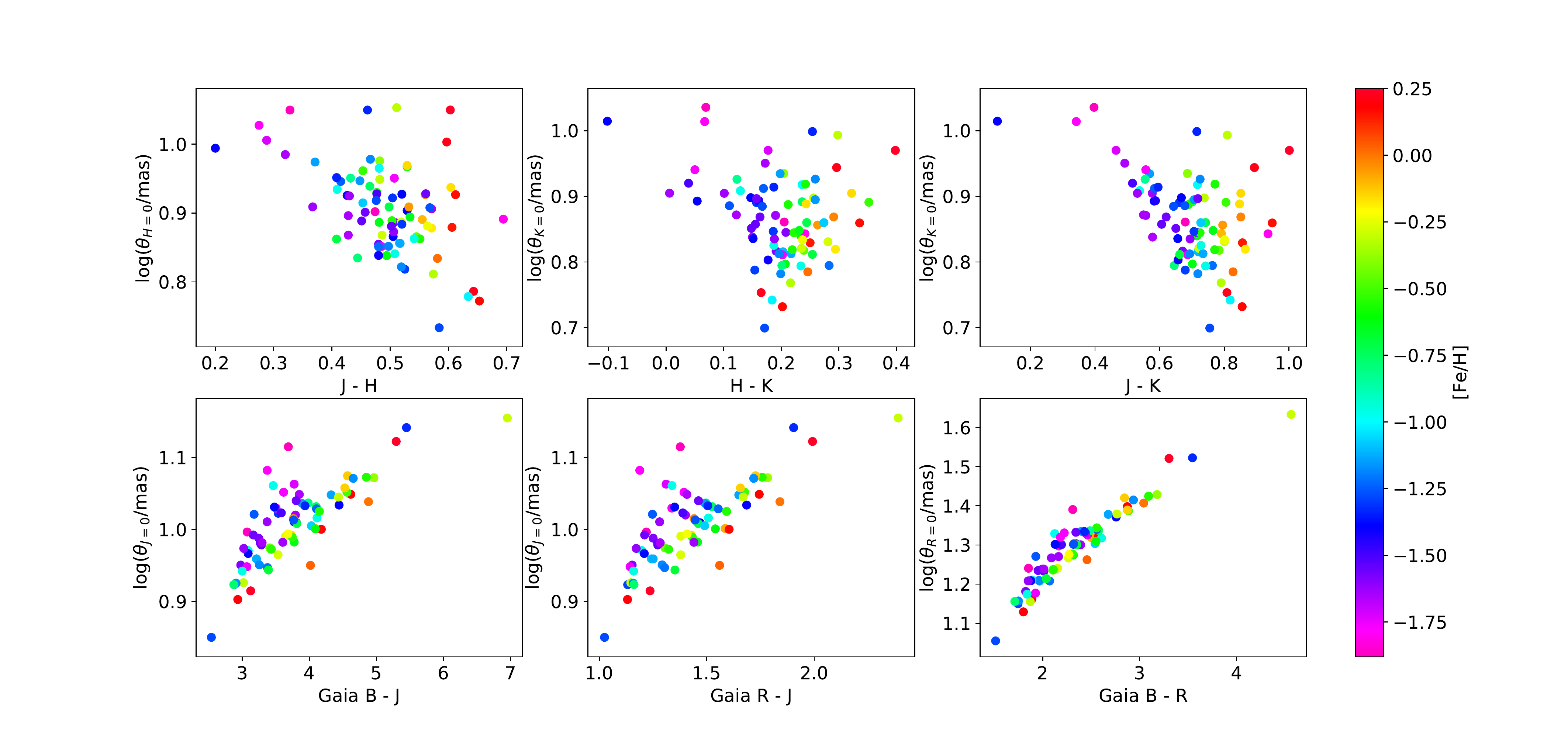}
\caption{\small
Angular diameter at zero apparent magnitude versus color for each of the stars in our sample. We have chosen the filters displayed in this plot because they are similar to the filters that will be available on \textit{WFIRST}. The points are colored by our estimated metallicities, and as expected we find that for a given color the angular diameter changes with metallicity. The tightest relation (least scatter in angular diameter for a given color) is given from the Gaia B - Gaia R bands. }
\label{f:angDiam}
\end{center}
\end{figure*}

We have used the radii to derive relations for angular diameter versus color, which will be useful for \textit{WFIRST}'s exoplanet microlensing survey (as discussed in Section \ref{s:intro}). Figure \ref{f:angDiam} shows the angular diameter of our sample at zero apparent magnitude in different filters ($\theta_{m = 0}$) versus color. $\theta_{m = 0}$ is proportional to surface brightness and is used in constraining exoplanet properties from microlensing events. 

\begin{figure*}[ht]
\begin{center}
\includegraphics[width=\linewidth]{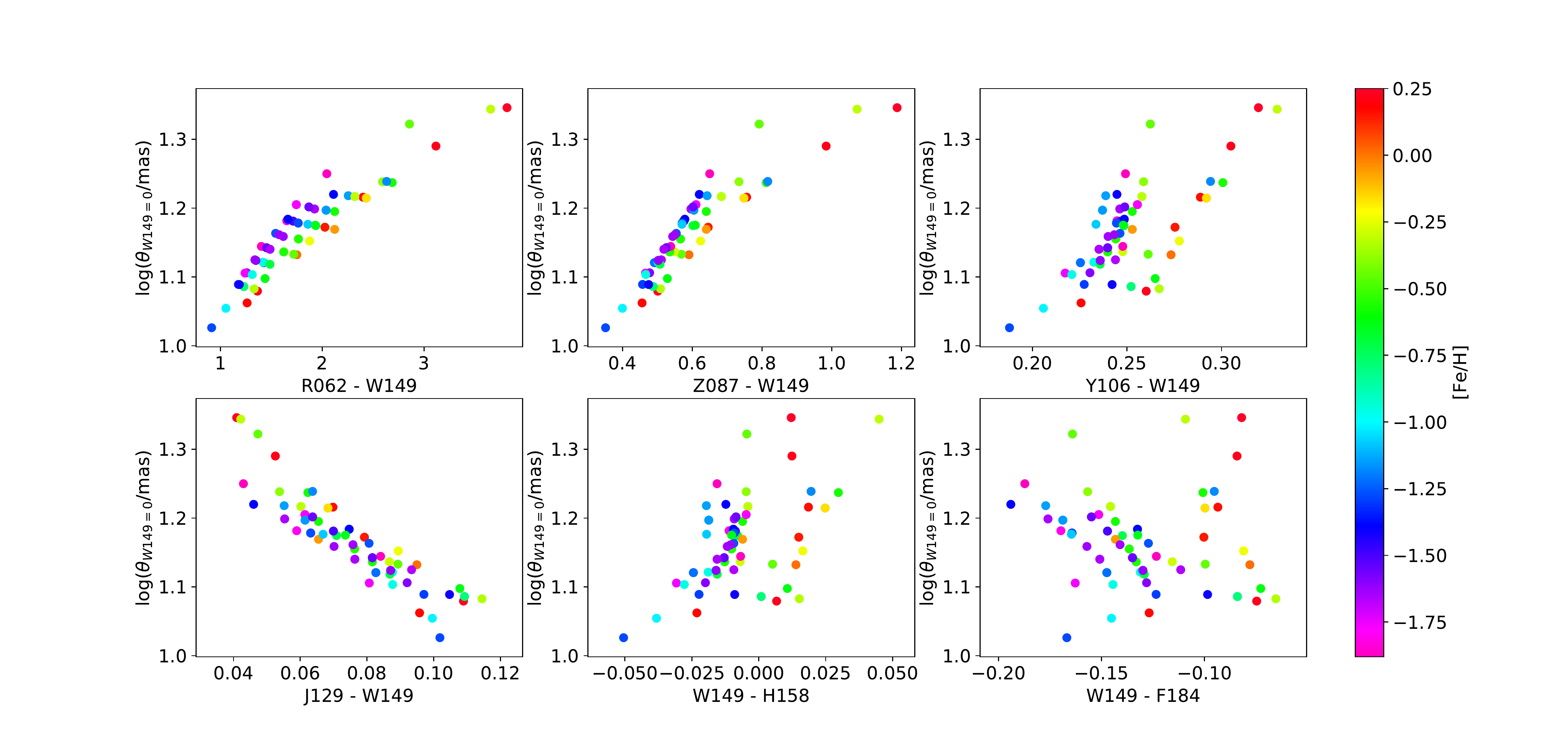}
\caption{\small
Angular diameter at zero apparent magnitude versus synthetic color for each of the stars in our sample. The synthetic photometry shows very similar trends as seen in Figure \ref{f:angDiam}. We also choose to show the W149 filter since this filter will be used for the detection of microlensing events. We plot all the other filters planned for \textit{WFIRST} to find the second filter that will reduce the uncertainty in the angular diameter for a given color due to the differing values of [Fe/H]. The filters are all labeled by their band name, and then a number which gives the central wavelength of the filter in units of $10^{-8} m$ (i.e. Z087 has a central wavelength of $0.87 \mu m$).  }
\label{f:angDiamWFIRST}
\end{center}
\end{figure*}

We also present similar relations using synthetic photometry for the proposed \textit{WFIRST} filters (Figure \ref{f:angDiamWFIRST}). The wide near-IR band (W149) ranges from approximately $1 - 2 \mu m$, and will be used to detect microlensing events. We test colors containing W149 and the six other proposed filters to see which color combination has the smallest change in angular diameter for a given color. We find that the $z$-band filter (Z087) reduces the uncertainty in the angular diameter the most, but there is still a clear metallicity trend present. The fractional uncertainty on the host and planet mass is equal to the fractional uncertainty in $\theta_{m = 0}$ ($\sigma_{\theta} / \theta$). We find that the fractional uncertainty in $\theta_{m = 0}$ is 5\%, and for some of the cooler stars can be as high as 12\%. For comparison, a fractional precision of $\sim 7\%$ is achievable with ground-based microlensing data sets for blue stars using optical filters, where the uncertainty is dominated by dereddening and not the angular diameter-color relations (see Section 4.3 of \citealt{gould15}  and \citealt{gould14} for a detailed discussion).

\begin{figure*}[ht]
\begin{center}
\includegraphics[width=\linewidth]{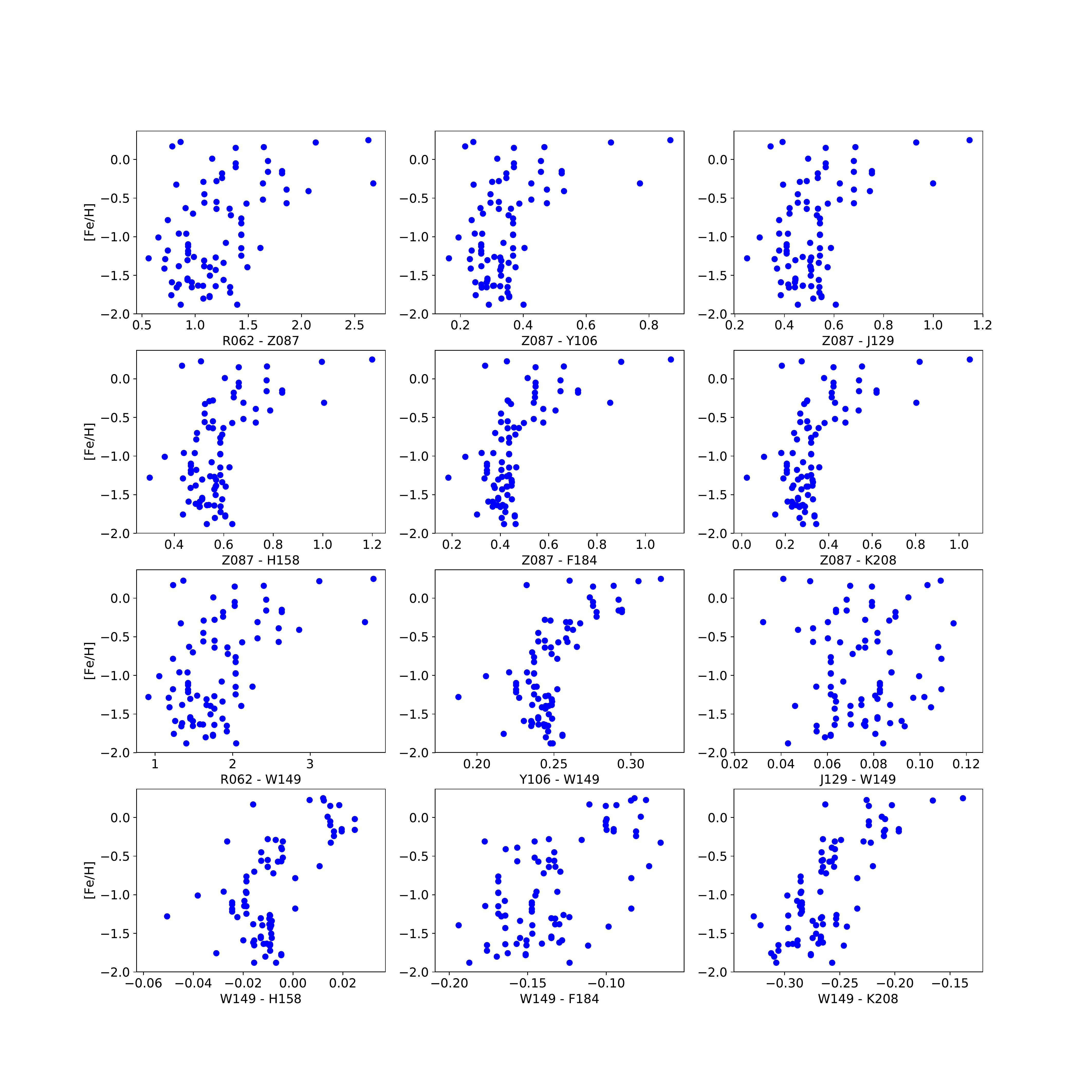}
\caption{\small
\textit{WFIRST} synthetic colors versus our estimated metallicities for all the stars in our sample. We test every combination that includes either the Z087 or the W149 filter to find the best third filter to break the metallicity color degeneracy present in Figure \ref{f:angDiamWFIRST}. An ideal color$-$metallicity relation will be linear for the entire metallicity regime, and show a small scatter around the linear relation. All of the color relations that contain the Z087 filter seem to saturate below about $-$1.0 dex, and a decrease in metallicity no longer corresponds to a change in color. The W149 - K208 color relation shows the least amount of scatter ($1-\sigma$ uncertainty is 0.4 dex). The W149-F184 color relation has slightly more scatter ($1-\sigma$ uncertainty is 0.5 dex), but still shows a linear trend. The appearance of outliers on the right of the W149-F184 relation that form a ``second-sequence" is due to the coarse grid resolution publicly available for the BT Settl models. We find that all the points in this ``second-sequence'' have a log($g$) of 4.5 dex whereas the majority of the rest of the targets have best-fit log($g$) values of 5.0 or 5.5. With a smoother grid resolution (or real data) we suspect that these outliers would disappear. }
\label{f:WFIRSTmetal}
\end{center}
\end{figure*}

The degeneracy between color and metallicity can be broken with the addition of a third filter, which can be used to estimate the metallicity of the source star and in turn obtain a more accurate estimate of the source star's angular diameter. We test all the different filter combinations that contain either the W149 filter or the Z087 filter (see Figure \ref{f:WFIRSTmetal}). 
The color combination that gives the smallest uncertainty in the metallicity (0.4 dex) is the K208 and W149 filters. This filter combination also shows a linear trend with metallicity throughout our metallicity range, but it only has a dynamic range of $\sim 0.15$ magnitudes. The Z087-F184 relation comparatively has a dynamic range of $\sim 0.75$ magnitudes, while still having a tight relation (uncertainty of 0.52 dex). If bulge stars below $-$1.0 dex are determined to be rare, we can use the Z087-F184 to get metallicities for stars above $\sim -$1.0 dex. 
However, if the probability of observing M subdwarfs of even lower metallicity ([Fe/H] $< -1.0$ dex) is determined to be substantial, the most linear relation, W149-K208, would provide the best discrimination across a wider range of [Fe/H]. K208 is not currently included in WFIRST's filter wheel, but has been considered in the past, and may be included in the future.

By adding in a third filter the scatter in the log of the angular diameter can be reduced to $3\%$ (from about $5\%$). We conclude that while adding a third filter will reduce our fractional uncertainty, without a third filter the results are still promising that we can obtain accurate angular diameters for the vast majority of targets. 

We remind the reader that this section makes use of synthetic photometry, generated using model atmospheres, which have been shown to have discrepancies in the M dwarf regime. While we believe all of the overall trends shown by the synthetic photometry, individual values may differ by small amounts. We also note that the filter profiles used here are not necessarily the ones that will end up on the \textit{WFIRST} mission and so more testing will be required at a later stage in the mission planning.  

\section{Discussion}

\subsection{Internal Consistency Check}

\begin{figure*}[ht]
\begin{center}
\includegraphics[scale=0.5]{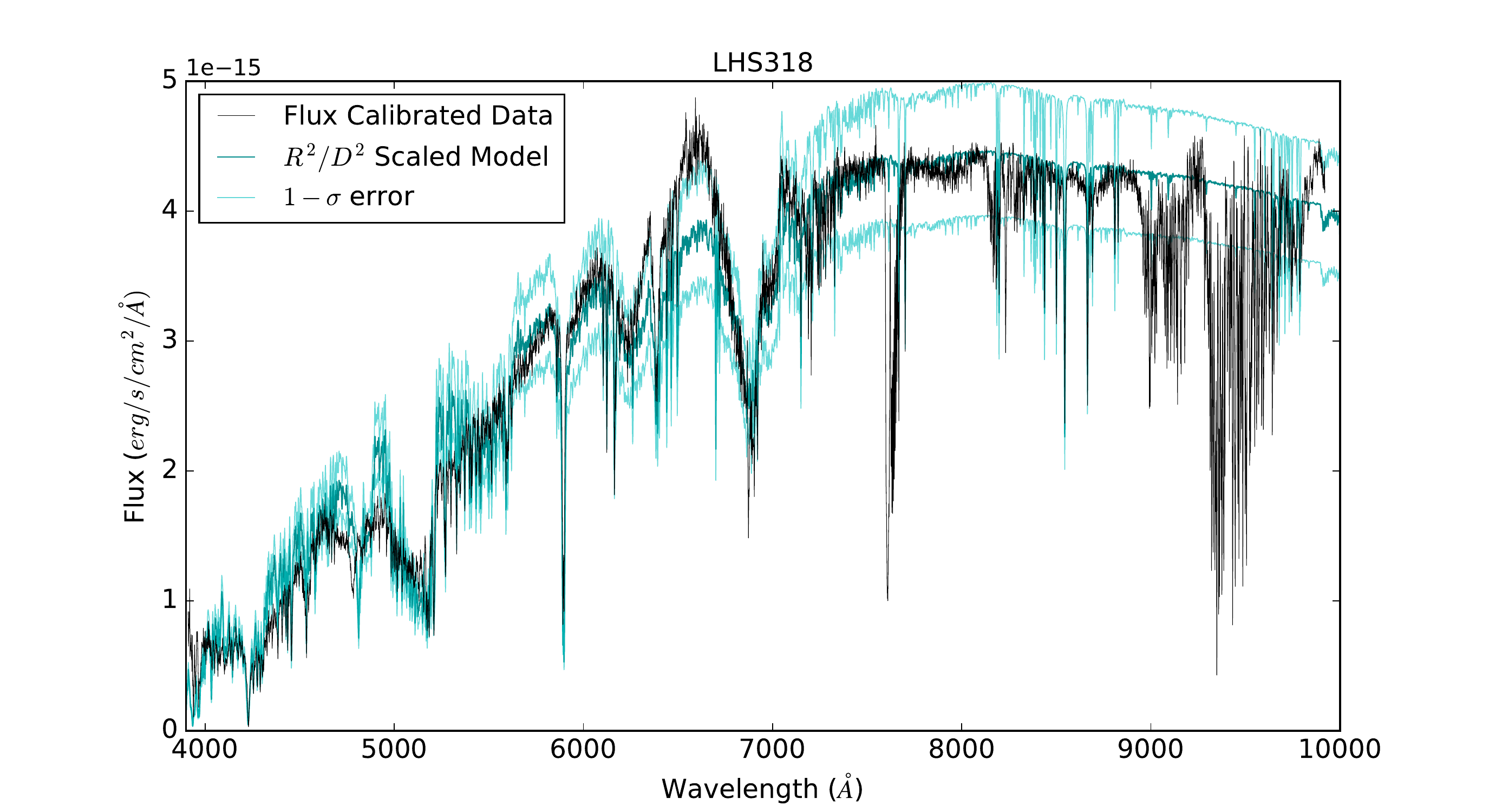}
\caption{\small
An example of our internal consistency check. The flux calibrated observed spectrum is plotted in black, while the $R^2/D^2$ scaled model is plotted in teal. The two more transparent teal spectra show the model spectrum scaled using the $1-\sigma$ uncertainties on the radius. Note the large mismatch between the scaled model and the data around 7500 and between 9000 and 10000 \AA\ is due to telluric contamination in our spectra. }
\label{f:R2D2}
\end{center}
\end{figure*}

We can perform a self-consistency check on our radius determinations by comparing the apparent flux levels in our spectrum to the flux of the best fitting model, scaled by the dilution factor $R^2/D^2$ to determine the apparent flux from the model at Earth. Plotted in Figure \ref{f:R2D2} is an example of this consistency check. Any target where the observed flux calibrated (black) spectrum fell outside of the $R^2/D^2 \pm \sigma_{R^2/D^2}$  scaled model (transparent blue) was noted as being inconsistent. 

We find that 9 out of our 88 spectra fall outside of the $1-\sigma$ errorbars: Gl 436, Gl 447, Gl 51, LHS 170, LHS 375, LHS 2843, LHS 2852, LHS 3189, LHS 3255, LHS 3555. These 9 targets are some of the most extreme outliers in Figure \ref{f:TeffVsRad}, which suggests that the true scatter in the $T_\mathrm{eff}-$Radius relation is actually smaller than what is shown in Figure \ref{f:TeffVsRad}. We believe that the majority of this discrepancy is due to the uncertainty in $T_\mathrm{eff}$, and for the one source with previously determined parameters (Gl 436) this is the case; our $T_\mathrm{eff}$ estimate differs by $\sim$150 K from what \citet{mann15} report and thus our radius estimate differs by 0.06 $R_\mathrm{Sun}$. However, we hypothesize that the radius discrepancy in a few of our sources is due to inaccurate metallicities, which leads to poor fits to the models. LHS 170 is the hottest star in our sample, and for that reason may not have an accurate metallicity estimate since the methods used for determining metallicity for our sample are only valid for spectral types later than $\sim$K7. LHS 2852 has differing spectroscopic and photometric metallicities even though it is in a part of parameter space where both methods should be valid, leading us to conclude that there is potentially something odd about its metallicity. 

Because almost 90\% of our sources pass our internal consistency check we are confident that the overall trends observed in our data are accurate. We can also conclude that our $1-\sigma$ errorbars do not seem to be underestimated, and if anything they are overestimated. 

\subsection{Variations in Chemical Abundances}

\begin{figure*}[ht]
\begin{center}
\includegraphics[scale=0.8]{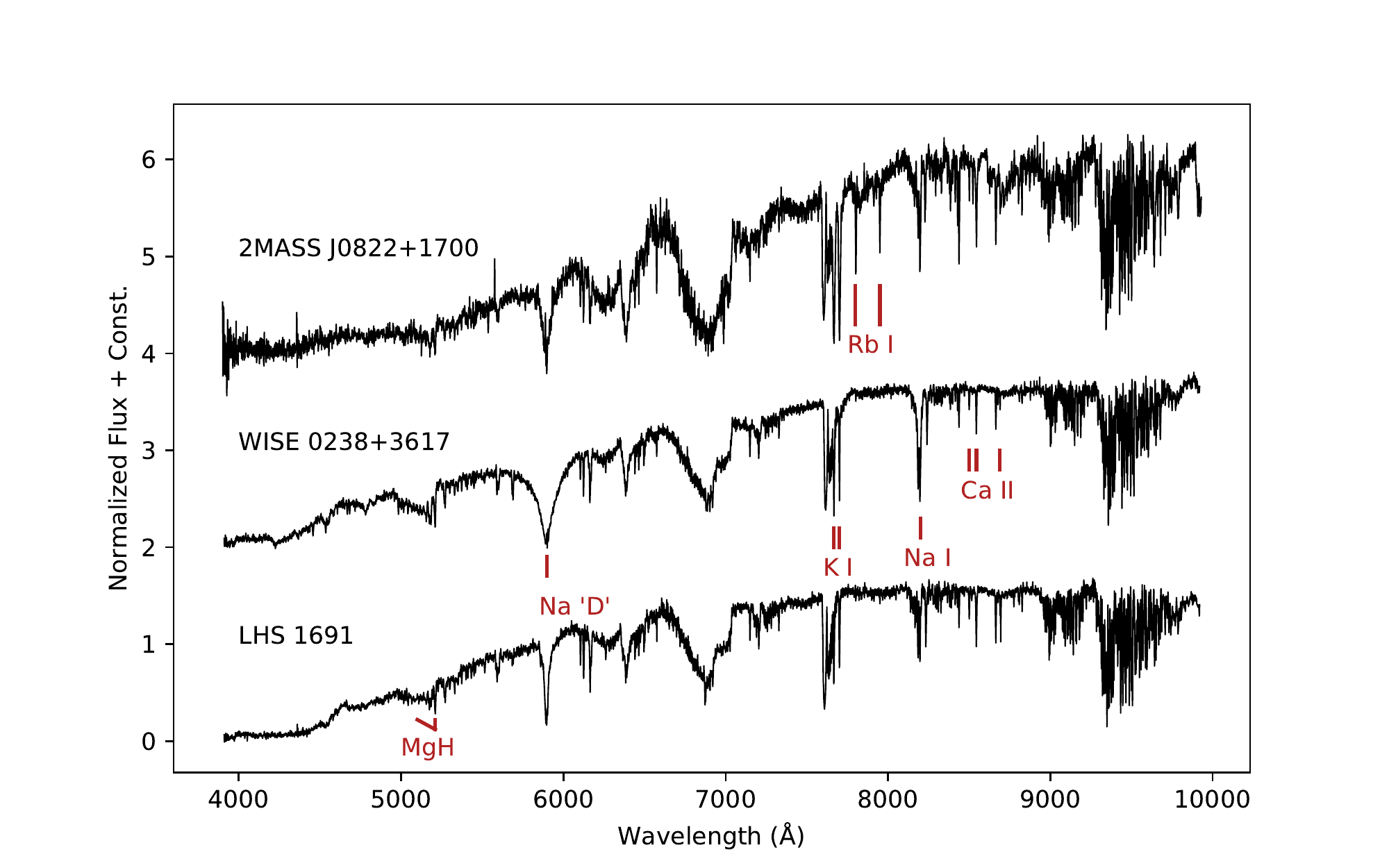}
\caption{\small
Figure showing three of our spectra that have peculiar features. 2MASS J0822+1700 has Rb I lines that are not seen in any other spectra. WISE 0238+3617 has an extremely broad Na `D' doublet, slightly stronger Na I and K I, as well as weaker Ca II lines. LHS 1691 has a particularly weak MgH band and the bluest end of the spectrum is noticeably smoother than spectra of similar spectral type.  }
\label{f:odd}
\end{center}
\end{figure*}

Many of our spectra have unusual spectral features that are not reproduced by the stellar atmosphere models, or have colors and spectroscopic metallicities that are at odds. Figure \ref{f:odd} shows these spectra with the features in question labeled. 
2MASS J0822+1700 contains prominent Rb I lines (first noted in \citet{lepine04}), which are not seen in any other spectra in our sample or in the models. Rb is a slow neutron capture (s-process) element formed during the AGB phase of stellar evolution, so these interesting objects could have formed near an AGB star and hence be polluted by an overabundance of Rb compared to [Fe/H]. This effect has been seen in warmer halo stars that exhibit enhancements in s-process elements \citep{beers05}. 

WISE 0238+3617 has a significantly broader Na doublet (labeled Na `D' in Figure \ref{f:odd}) than any of our other spectra, as well as a deeper Na I doublet ($\sim 8200$ \AA), deeper K I lines, and weaker Ca II lines. \citet{kirkpatrick16}, who first published its spectrum, theorized that the broad Na doublet was indicative of an extremely low metallicity ($<-2.0$ dex). The extremely broad Na doublet could be indicative of an over-enhancement of Na. Na is produced during C burning in SN II, so this star could have environmental enhancement, but more information is needed to verify this claim.  

LHS 1691 has weak absorption from the MgH band compared to other spectra of similar spectral type. Evidence for two populations of metal poor stars with different Mg abundances (low- and high-Mg groups) has been seen by many groups in the halo population \citep[e.g., ][]{hayes18}. We hypothesize that the weak MgH absorption in LHS 1691 suggests that this star is part of a low-Mg population. There are other stars in our sample with varying strengths in MgH for similar spectral types, which could be indicative of the spread in the [Mg/Fe] versus [Fe/H] measured by \citet{hayes18} (see Figure 3 in \citealt{hayes18}). Since Mg is an $\alpha$ element, publicly available models with varying $\alpha$ abundances for single [Fe/H] values would be useful to better model subdwarfs and estimate $\alpha$ abundances for different stars. 

Some of our spectra also have spectroscopic features that are reminiscent of subdwarfs (very little TiO absorption), but near-IR colors that would point towards a dwarf star metallicity when the relation from \citet{newton14} is applied. LHS 1691 is the most extreme of these cases, where spectroscopically it is classified as an ultra subdwarf ($-$1.8 dex), but the photometric metallicity relation estimates a metallicity of +0.3 dex. Other stars that exhibit this behavior but are not as extreme are: WISE 0238+3617, LHS 2843, LHS 3382, and LHS 104. We are unsure what causes this interesting effect and merely note it in this paper, to be further explored at a later time. 

All of the above-mentioned unusual spectral features lead us to conclude that a single metallicity value with corresponding $\alpha$ abundance cannot always reproduce observe features, and that in reality the chemical composition of the stars in our sample is more complex.

\section{Conclusions}

We find that for a given temperature, an ultra subdwarf can be smaller than a dwarf star by up to a factor of five, and that the \citet{baraffe97} stellar evolution models are in agreement with our data, providing some of the first validation of these models for the lowest stellar temperatures and metallicities.  We also present relations that can be used to convert direct observables, such as color and absolute K-band magnitude, to stellar radii for metallicities down to $-$2.0 dex with radius uncertainties of $\sim 20\%$ and 6\%, respectively.

Finally, we present color to absolute angular diameter relations that will be useful for the \textit{WFIRST} exoplanet microlensing survey. Many of the source stars observed by \textit{WFIRST} will be in the bulge of the galaxy, where metallicities range from $-$3.0 dex to +1.0 dex, and so the stellar angular diameters as a function of metallicity will be a required input to extract accurate exoplanet masses. We find that along with the W149 filter, the Z087 filter gives the least amount of scatter in the derived angular diameter due to metallicity change. However, the angular diameter of the source star can still change by $10-15\%$ due to a change in metallicity of 2.0 dex. To break this degeneracy a third filter can be used to estimate the metallicity. We find that the W149 - K208 color combination gives a linear color to metallicity relation with the smallest uncertainties.

\acknowledgements

The authors would like to thank the entire NEOWISE and CatWISE team at IPAC, as well as Avani Gowardhan, Rebecca Larson, Savannah Jacklin, Chris Theissen, and Pat Tamburo for useful conversations throughout the paper process. Part of this work was performed while AYK was a Visiting Graduate Student Research Fellow at the Infrared Processing and Analysis Center (IPAC), California Institute of Technology. 

This publication makes use of data products from the Two Micron All Sky Survey, which is a joint project of the University of Massachusetts and the Infrared Processing and Analysis Center/California Institute of Technology, funded by the National Aeronautics and Space Administration and the National Science Foundation. 

This publication makes use of data products from the
Wide-field Infrared Survey Explorer, which is a joint
project of the University of California, Los Angeles,
and the Jet Propulsion Laboratory/California Institute
of Technology, funded by the National Aeronautics and
Space Administration.

Funding for the Sloan Digital Sky Survey IV has been provided by the Alfred P. Sloan Foundation, the U.S. Department of Energy Office of Science, and the Participating Institutions. SDSS-IV acknowledges
support and resources from the Center for High-Performance Computing at
the University of Utah. The SDSS web site is www.sdss.org.

SDSS-IV is managed by the Astrophysical Research Consortium for the 
Participating Institutions of the SDSS Collaboration including the 
Brazilian Participation Group, the Carnegie Institution for Science, 
Carnegie Mellon University, the Chilean Participation Group, the French Participation Group, Harvard-Smithsonian Center for Astrophysics, 
Instituto de Astrof\'isica de Canarias, The Johns Hopkins University, 
Kavli Institute for the Physics and Mathematics of the Universe (IPMU) / 
University of Tokyo, the Korean Participation Group, Lawrence Berkeley National Laboratory, 
Leibniz Institut f\"ur Astrophysik Potsdam (AIP),  
Max-Planck-Institut f\"ur Astronomie (MPIA Heidelberg), 
Max-Planck-Institut f\"ur Astrophysik (MPA Garching), 
Max-Planck-Institut f\"ur Extraterrestrische Physik (MPE), 
National Astronomical Observatories of China, New Mexico State University, 
New York University, University of Notre Dame, 
Observat\'ario Nacional / MCTI, The Ohio State University, 
Pennsylvania State University, Shanghai Astronomical Observatory, 
United Kingdom Participation Group,
Universidad Nacional Aut\'onoma de M\'exico, University of Arizona, 
University of Colorado Boulder, University of Oxford, University of Portsmouth, 
University of Utah, University of Virginia, University of Washington, University of Wisconsin, 
Vanderbilt University, and Yale University.

The Pan-STARRS1 Surveys (PS1) and the PS1 public science archive have been made possible through contributions by the Institute for Astronomy, the University of Hawaii, the Pan-STARRS Project Office, the Max-Planck Society and its participating institutes, the Max Planck Institute for Astronomy, Heidelberg and the Max Planck Institute for Extraterrestrial Physics, Garching, The Johns Hopkins University, Durham University, the University of Edinburgh, the Queen's University Belfast, the Harvard-Smithsonian Center for Astrophysics, the Las Cumbres Observatory Global Telescope Network Incorporated, the National Central University of Taiwan, the Space Telescope Science Institute, the National Aeronautics and Space Administration under Grant No. NNX08AR22G issued through the Planetary Science Division of the NASA Science Mission Directorate, the National Science Foundation Grant No. AST-1238877, the University of Maryland, Eotvos Lorand University (ELTE), the Los Alamos National Laboratory, and the Gordon and Betty Moore Foundation.

This work made use of data from the European Space Agency (ESA)
mission {\it Gaia} (\url{https://www.cosmos.esa.int/gaia}), processed by
the {\it Gaia} Data Processing and Analysis Consortium (DPAC,
\url{https://www.cosmos.esa.int/web/gaia/dpac/consortium}). Funding
for the DPAC has been provided by national institutions, in particular
the institutions participating in the {\it Gaia} Multilateral Agreement.

This research has made use of the SVO Filter Profile Service (http://svo2.cab.inta-csic.es/theory/fps/) supported from the Spanish MINECO through grant AyA2014-55216. This research has benefitted from the SpeX Prism Library (and/or SpeX Prism Library Analysis Toolkit), maintained by Adam Burgasser at http://www.browndwarfs.org/spexprism. This research made use of Astropy, a community developed
core Python package for Astronomy \citep{astropy13}. This research
has made use of the SIMBAD database, operated at
CDS, Strasbourg, France. 

\facilities{IRTF (iSHELL), Palomar (DBSP)}

\software{Spextool \citep{cushing04}}

\clearpage
 

%

\end{document}